\documentclass[final,5p,times,twocolumn]{elsarticle}

\usepackage{amssymb}
\usepackage{amsthm}
\usepackage[switch]{lineno}

\usepackage{makecell}
\usepackage{longtable}
\usepackage{adjustbox}

\usepackage{enumitem}
\usepackage{multirow}
\usepackage{listings}
\usepackage{array}
\usepackage{xurl}
\newcolumntype{P}[1]{>{\centering\arraybackslash}p{#1}}
\newcolumntype{M}[1]{>{\centering\arraybackslash}m{#1}}

\usepackage{tikz}
\newcommand*\emptycirc[1][1ex]{\tikz\draw (0,0) circle (#1);}
\newcommand*\halfcirc[1][1ex]{
  \begin{tikzpicture}
  \draw[fill] (0,0)-- (90:#1) arc (90:270:#1) -- cycle;
  \draw (0,0) circle (#1);
  \end{tikzpicture}}
\newcommand*\fullcirc[1][1ex]{\tikz\fill (0,0) circle (#1);}

\journal{Energy \& Buildings}

\begin{document}

\begin{frontmatter}

\title{A Systematic Comparison and Evaluation of Building Ontologies for Deploying Data-Driven Analytics in Smart Buildings}

\author[a]{Zhangcheng Qiang\corref{cor}}
\ead{qzc438@gmail.com}
\author[b]{Stuart Hands}
\author[a]{Kerry Taylor}
\author[b]{Subbu Sethuvenkatraman}
\author[c]{Daniel Hugo}
\author[a]{Pouya Ghiasnezhad Omran}
\author[a]{Madhawa Perera}
\author[a]{Armin Haller}

\affiliation[a]{organization={Australian National University, School of Computing},
            addressline={108 North Road, Acton},
            city={Canberra},
            postcode={2601},
            state={ACT},
            country={Australia}}
            
\affiliation[b]{organization={CSIRO Newcastle Energy Centre},
            addressline={10 Murray Dwyer Circuit, Mayfield West},
            city={Newcastle},
            postcode={2304},
            state={NSW},
            country={Australia}}

\affiliation[c]{organization={CSIRO Data61},
            addressline={College Road, Sandy Bay},
            city={Hobart},
            postcode={7005},
            state={TAS},
            country={Australia}}

\cortext[cor]{Corresponding author.}

\begin{abstract}

Ontologies play a critical role in data exchange, information integration, and knowledge sharing across diverse smart building applications. Yet, semantic differences between the prevailing building ontologies hamper their purpose of bringing data interoperability and restrict the ability to reuse building ontologies in real-world applications. In this paper, we propose and adopt a framework to conduct a systematic comparison and evaluation of four popular building ontologies (Brick Schema, RealEstateCore, Project Haystack and Google's Digital Buildings) from both axiomatic design and assertions in a use case, namely the Terminological Box (TBox) evaluation and the Assertion Box (ABox) evaluation. In the TBox evaluation, we use the SQuaRE-based Ontology Quality Evaluation (OQuaRE) Framework and concede that Project Haystack and Brick Schema are more compact with respect to the ontology axiomatic design. In the ABox evaluation, we apply an empirical study with sample building data that suggests that Brick Schema and RealEstateCore have greater completeness and expressiveness in capturing the main concepts and relations within the building domain. The results implicitly indicate that there is no universal building ontology for integrating Linked Building Data (LBD). We discuss ontology compatibility and investigate building ontology design patterns (ODPs) to support ontology matching, alignment, and harmonisation.

\end{abstract}


\end{frontmatter}

\section{Introduction}
\label{sec: introduction}

Smart buildings, also known as digital buildings or intelligent buildings, have received significant attention in recent years and are contributing to digital innovation in the building industry. The global smart building market is envisioned to grow at a compound annual growth rate (CAGR) of 10.9\% by 2026, with a total market value of USD 121.6 billion~\cite{smartbuildingmarket2021}. The popularity of smart buildings is due to their ability to provide effective and efficient management of building facilities and optimum usage of resources through intelligent building systems~\cite{pritoni2021metadata}.

A smart building is a complex concatenation of structures, systems, and technology. Smart building data shows typical characteristics of ``big data'' viz: high volume, variety, velocity, veracity, and value (5Vs)~\cite{bashir2017iot}. A smart building can easily collect thousands of data points per hour. This data can come from a wide range of resources, including smart equipment, building IoT platforms, building management systems (BMS), and building facility and maintenance systems~\cite{miller2017building}. Currently, there are different approaches used by service providers and vendors for managing building data. While some buildings may have customised data models to describe relevant context, functions, and interactions within buildings, oftentimes, buildings have limited information available to represent data along with the context~\cite{smartbuildingsurvey2019}. Various studies have identified the importance of a uniform data model that could exchange building data across various sensors, devices, systems, and buildings in an integrated, dynamic, and functional manner~\cite{venugopal2012semantics,fierro2022survey,lbd2023}.

Ontologies are human-readable and machine-interpretable representations that classify entities that may exist and the relationships between them, enabling richer data description, information integration, and knowledge sharing~\cite{stephan2007knowledge}. Notably, a recent review of knowledge management in the building industry~\cite{deng2022transforming} shows that ontologies have great research significance in academia and the most extensive application fields, including the deployment of data-driven digital construction~\cite{kukkonen2022ontology,zheng2021shared}, predictive analytics and diagnostics~\cite{hu2021building,wang2021ontology,ren2021aligning}, and automated system optimisation (ASO)~\cite{schneider2020design,cao2022ontology,li2022ontology}. Approximately 90\% of the building management systems developed in the past ten years are built on (or are adopting) domain ontologies~\cite{luo2021overview}. Among these building ontologies, Brick Schema~\cite{balaji2016brick} (abbr. ``Brick''), RealEstateCore~\cite{hammar2019realestatecore} (abbr. ``RECore''), Project Haystack~\cite{john2020project} (abbr. ``Haystack''), and Google's Digital Buildings~\cite{berkoben2020digital} (abbr. ``DB'') are some of the well-known ontologies that are related to operational aspects of buildings including their air conditioning and electrical systems. See Section~\ref{subsec: selected building ontologies} and~\ref{subsec: selected building concepts} for the reasons these particular building ontologies and concepts were selected for this study. While these building ontologies tend to capture similar details and contextual information in a building, they are developed and maintained by different organisations for different purposes. As a result, building ontologies may differ in the level of abstraction. The overlapping nature is often confusing for the users, and different ways of expressing entities and their relationships hamper the interoperability between these ontologies.

The existence of different ontologies to represent the same type of building systems presents many challenges for semantic model developers and users. They are often unsure of which ontology to use to create building data models. The scalability of data-driven analytics is severely hampered due to existing interoperability limitations. While each ontology has evolved to meet a specific set of needs and abstract building information at multiple levels, there is no consistent and interpretable semantics to verify their former relationships~\cite{ASHRAE223P}, nor clearly defined case boundaries for each building ontology~\cite{brickrecore2022}. Mixing ontologies, though permitted in RDF~\cite{rdf2014}, may be impractical due to differing verification rules (e.g., OWL~\cite{owl2012} vs SHACL~\cite{shacl2017}) or query options (e.g., matching types vs tags).

This paper conducts a systematic comparison and evaluation of four building ontologies both in the Terminological Box (TBox) and the Assertion Box (ABox)\footnotemark. We discover ontology compatibility and investigate their design patterns, aiming to support building ontology matching, alignment, and harmonisation. In the TBox evaluation, we employ the SQuaRE-based Ontology Quality Evaluation (OQuaRE) framework~\cite{duque2011oquare} to analyse the axiomatic design of building ontologies. In the ABox evaluation, we conduct an empirical case study of a smart office building located in Melbourne with a real-world application-level use case of the energy audit. Specifically, this paper makes the following contributions:

\footnotetext{There are many ontology TBox and ABox evaluation methods available for use. There are no universal or optimal methods, and results are subject to the context of each case. One of our purposes is to introduce a generic framework for ontology comparison and selection. A comprehensive evaluation of TBox and ABox evaluation methods is out of scope here.}

\begin{itemize}[wide, itemsep=0pt, topsep=0pt, labelindent=0pt]
\item A Terminological Box (TBox) evaluation of four popular building ontologies using the OQuaRE Framework. Characteristics examined include Structural, Functional Adequacy, Maintainability, Transferability, Reliability, Compatibility, and Operability.
\item An Assertion Box (ABox) evaluation of four popular building ontologies from an empirical study of modelling a medium-sized office building in Melbourne, Australia. HVAC (Heating, Ventilation, and Air Conditioning) related Building Spaces, Building Equipment and Systems, Building Points and Measurements have been considered in the study.
\item A systemic analysis of compatibility between ontologies, and the common and conflicting design patterns in these building ontologies. We also present our preliminary results of Building Ontology Design Patterns (ODPs).
\end{itemize}

The remainder of the paper is structured as follows. Section~\ref{sec: related work} reviews the related work, and Section~\ref{sec: methodology} demonstrates the methodology used in this paper. We present the detailed comparison of four building ontologies in Section~\ref{sec: results} (TBox in Section~\ref{sec: tbox} and ABox in Section~\ref{sec: abox}, respectively), followed by a discussion in Section~\ref{sec: discussion}. The preliminary results of building ODPs are proposed in Section~\ref{sec: building ODPs}. Section~\ref{sec: conclusions} concludes the paper.

\section{Related Work}
\label{sec: related work}

\subsection{Preliminary Notions: The TBox and ABox Evaluation}

Ontologists have distinguished two types of statements in ontologies for decades, commonly referred to as the Terminological Box (TBox) and the Assertion Box (Abox)~\cite{gruber1993translation}. The TBox is a set of ``schema'' axioms (classes and their definitions). For example, an elevator is defined as ``A device that provides vertical transportation between floors, levels or decks of a building, vessel or other structure''. In contrast, the ABox is a set of ``data'' axioms. They contain instance information about actual individuals and how these individuals are linked. ABox examples included statements like ``Elevator E1 is manufactured by Kone''. A Knowledge Base (KB) is a TBox plus an ABox.

The TBox evaluation could provide a superficial view of the structural design of a given ontology. In contrast, the ABox evaluation could provide application-level detailed information and explore the key concepts and relationships. Ontologists can apply different world views (i.e., TBox) to the same ground facts (i.e., ABox).

\subsection{Building Ontology Comparison and Evaluation}

The majority of works comparing building ontologies are conducted horizontally, i.e., comparing the same building ontologies in different applications. For example, the evaluation of Brick is based on eight canonical applications in six diverse buildings. Brick is expressive enough to capture 98\% of BMS data points on average in a wide variety of contexts~\cite{balaji2016brick}. Similarly, RECore evaluates the ontology by applying real-world applications to three real estate companies and one research project. A pilot deployment at property company Vasakronan shows RECore could cover in excess of 80,000 physical sensors and actuators and 20,000 external signals in ten buildings~\cite{hammar2019realestatecore}.

There are some attempts to conduct the comparison vertically, i.e., comparing different building ontologies on the same data. For example, authors in~\cite{bhattacharya2015short} evaluate Haystack, Industry Foundation Classes (IFC)~\cite{ifc2018}, and Semantic Sensor Network (SSN)~\cite{compton2012ssn} in three testbed buildings and quantify their shortcomings in completeness, expressiveness, and flexibility. In a different study~\cite{quinn2021case}, authors apply two industry-recognised ontologies, Brick and Haystack, to a large industry dataset called KGS Clockwork. The result shows that Brick is the superior ontology to exhibit HVAC concepts for smart buildings.

The TBox evaluation, however, is not included in these works. The ABox evaluation conducted in these works is at a horizontal level, or a vertical evaluation that only includes one or two ontologies. This means it is infeasible to abstract key concepts and relationships, and investigate common and conflicting design patterns. In this paper, we apply the analysis to both TBox and ABox. Also, we compare the ABox in four well-known building ontologies at a vertical level, i.e., applying one building sample data to different building ontologies.

\section{Methodology}
\label{sec: methodology}

\begin{figure}[!hbt]
\centering
\includegraphics[width=1\linewidth]{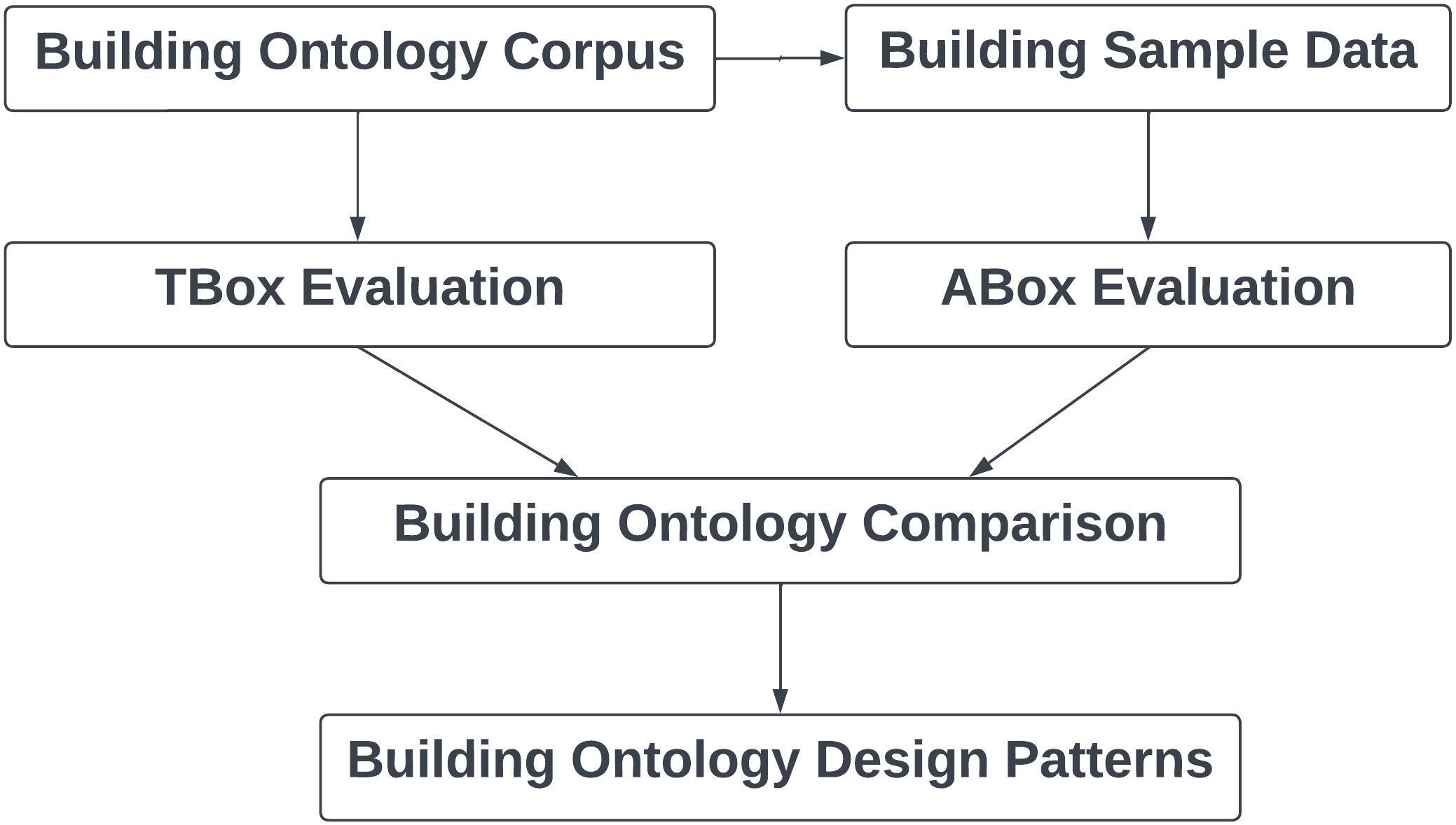}
\caption{Methodology used in this paper.}
\label{fig: methodology}
\end{figure}

Fig.~\ref{fig: methodology} shows the methodology used in this paper. We first select the building ontology corpus, and then the comparison is conducted in two phases: the TBox evaluation and the ABox evaluation. We apply four different building ontologies (i.e., TBox) to the same building sample (i.e., ABox). The TBox evaluation focuses on evaluating ontology quality (axiomatic design), while the ABox evaluation emphasises concept coverage and information retrieval in real-world use cases (assertions). Finally, we systematically compare four building ontologies with respect to ontology compatibility, and propose generic building ontology design patterns (ODPs).

\subsection{Selected Building Ontologies}
\label{subsec: selected building ontologies}

In this paper, we use Brick 1.2, RECore 3.3, Haystack 3.9.11 (referred to as Version 4 in marketing material), and DB 0.0.1. The online links of the selected building ontologies are listed in Table~\ref{tab: ontologies and namespaces}. We examined 30 building-related ontologies and chose these four ontologies. They cover the majority of the building concepts and are under continuous development\footnotemark.

\footnotetext{The complete list can be found in~\ref{appendix: building ontology corpus}.}

\begin{table}[!hbt]
\caption{Selected building ontologies and their online links.}
\label{tab: ontologies and namespaces}
\begin{center}
\begin{adjustbox}{width=1\columnwidth,center}
\begin{tabular}{|l|l|}
\hline
Ontologies      & Online Links\\
\hline
\small Brick    & \small \url{https://brickschema.org/}\\
\small RECore   & \small \url{https://www.realestatecore.io/}\\
\small Haystack & \small \url{https://project-haystack.org/}\\
\small DB       & \small \url{https://google.github.io/digitalbuildings/}\\
\hline
\end{tabular}
\end{adjustbox}
\end{center}
\end{table}

\subsection{Selected Building Concepts}
\label{subsec: selected building concepts}

\begin{figure}[!hbt]
\centering
\includegraphics[width=0.9\linewidth]{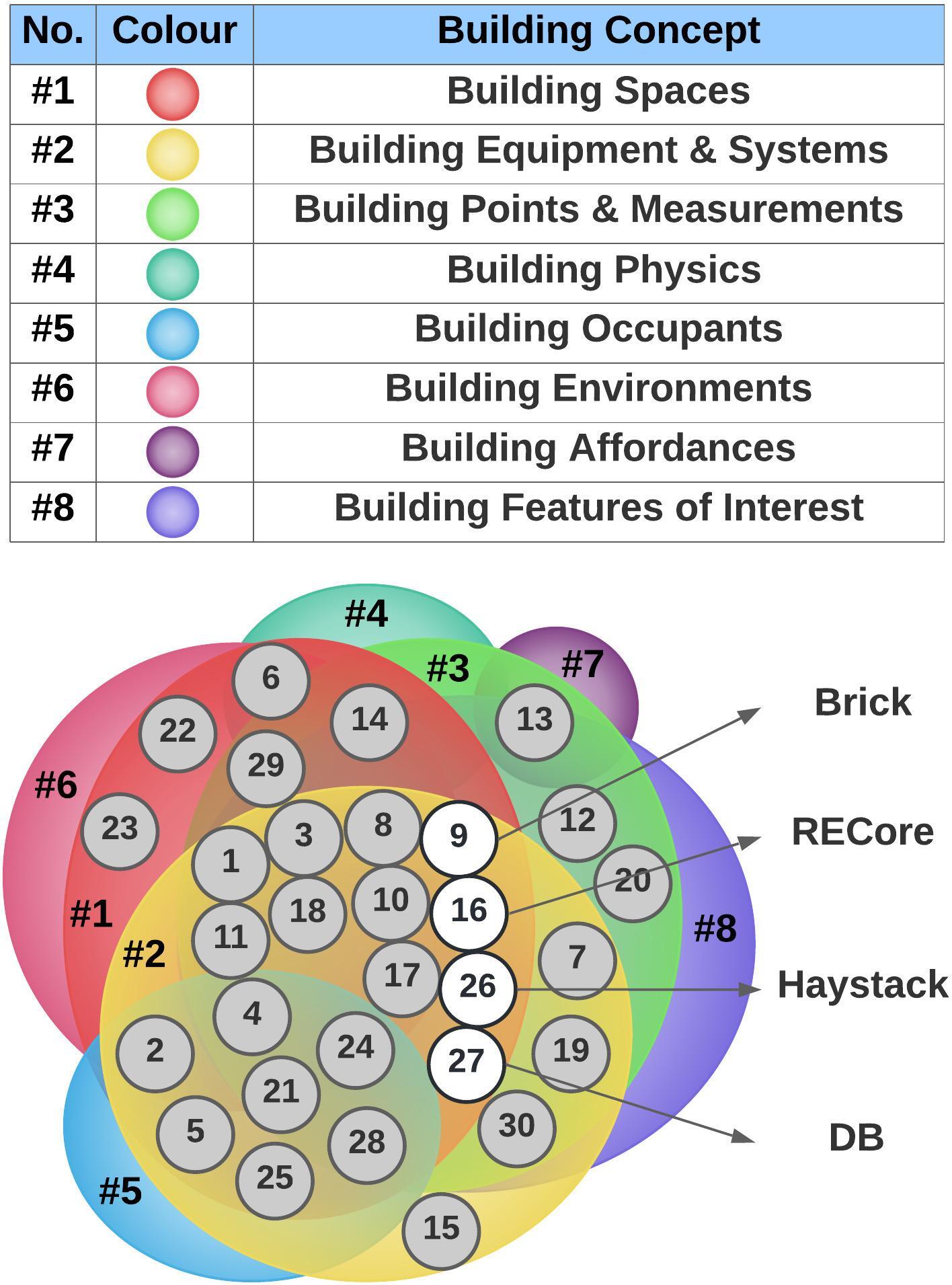}
\caption{Building concepts in 30 building and related domain ontologies.}
\label{fig: building concepts}
\end{figure}

The building concepts include three main parts: (1) Building Spaces, (2) Building Equipment and Systems, and (3) Building Points and Measurements. We examined the concepts defined in 30 building ontologies and building-related domain ontologies (shown in Fig.~\ref{fig: building concepts}), and these three are the top-ranked concepts that most of the ontologies include and that are closely related to smart building applications.

\subsection{The TBox Evaluation: The OQuaRE Framework}

The notion of SQuaRE-based Ontology Quality Evaluation (OQuaRE) is inspired by Software product Quality Requirements and Evaluation (SQuaRE), an evaluation standard for software products~\cite{zubrow2004software}. As ontologies can be viewed as software artifacts of the application of a conceptual modelling process, authors in~\cite{duque2011oquare} refine this standard to evaluate the ontology quality based on the graph theory. 

\begin{figure}[!hbt]
\centering
\includegraphics[width=1\linewidth]{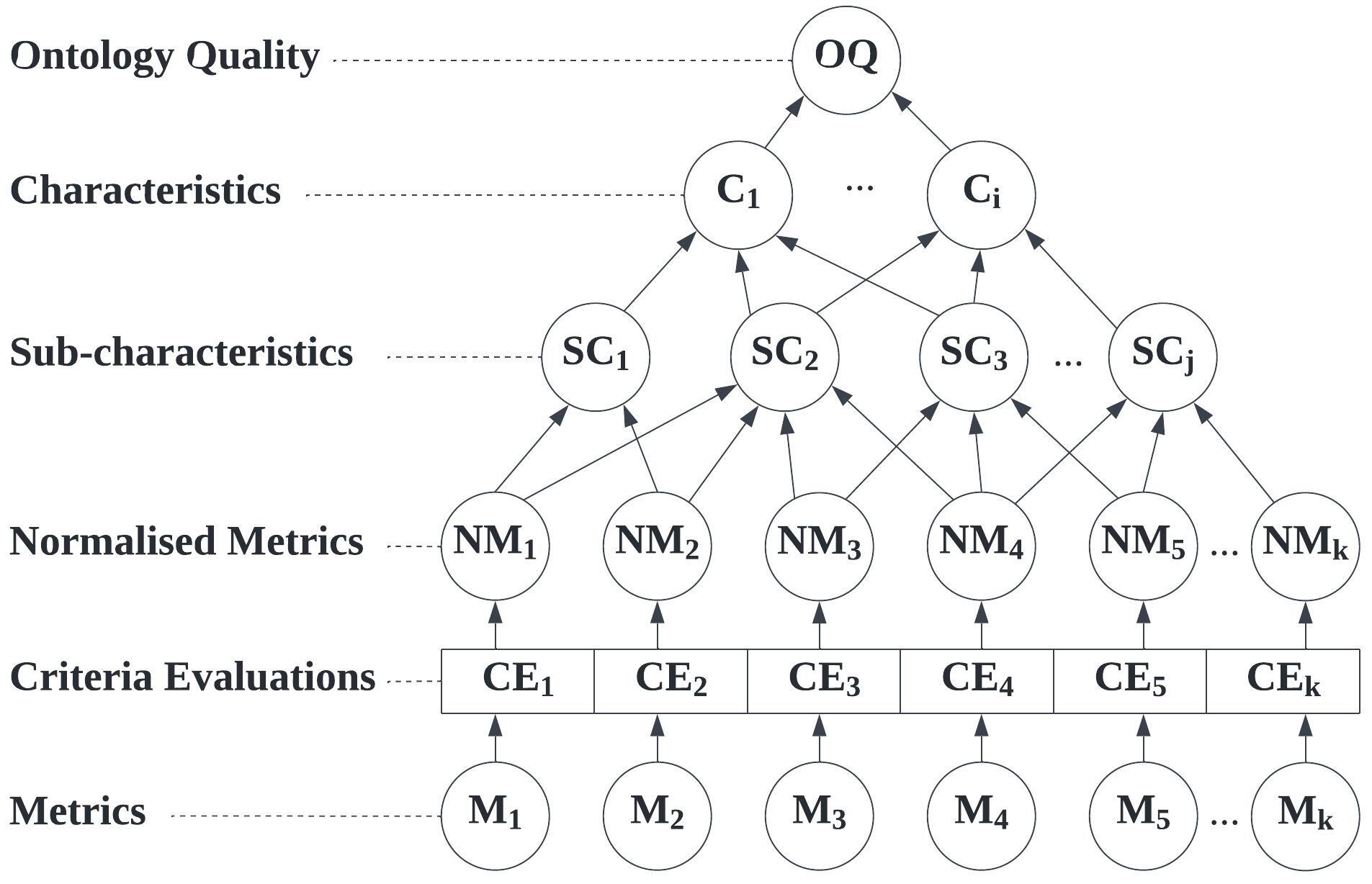}
\caption{The architecture of the OQuaRE Framework.}
\label{fig: OQuaRE}
\end{figure}

The OQuaRE framework is a multi-layer ontology quality evaluation model that follows a bottom-up approach (shown in Fig.~\ref{fig: OQuaRE}). The overall ontology quality is determined by a set of characteristics (C$_i$) and sub-characteristics (SC$_j$) with a family of metrics (M$_k$). As the value range of each metric is variant, we use a criteria evaluation (CE$_k$) to normalise a 5-scale quality score (NM$_k$). Then, these scores are categorised into different characteristics and sub-characteristics to calculate the mean value. The overall ontology quality score is the average quality score of its characteristics and sub-characteristics.

In this paper, we employ the OQuaRE framework as part of our evaluation process. OQuaRE has been applied in a wide range of domains~\cite{mozzaquatro2018ontology,alghamdi2019quantitative,roldan2020ontology,munoz2021gente,cornejo2021ontoslam,mishra2022evaluating,tibaut2022framework,sanchez2023ontology} and brings significant benefits. Firstly, it tends to convert the ontology TBox evaluation into an engineering activity that follows the guidelines of an ISO standard (ISO/IEC 25000:2005). A series of calculated quantitative values can be generated during the evaluation process. Secondly, the OQuaRE framework is semi-automatic and requires few domain experts to be involved, and only to select and evolve characteristics and sub-characteristics at the early stage. Evaluations in~\cite{duque2013evaluation} claim that there is no statistical difference between the results automatically generated by the OQuaRE framework and the manual evaluations carried out by domain experts. Lastly, OQuaRE is a generic framework that can be tailored to each particular domain by giving priorities, or changing and evolving characteristics and sub-characteristics. To the best of our knowledge, this is the first attempt to use the OQuaRE framework to evaluate building ontologies. In our use case, we use the framework to develop a highly-scalable, domain-specific, semi-automatic TBox comparison and evaluation for building ontologies.

\subsection{The ABox Evaluation: An Empirical Study}

Our comparison is based on different building ontologies representing the same sample building data, retrieved from a medium-sized office building in Melbourne. The building has 3 storeys, 69 rooms, and 18 HVAC zones. The building data is collected from a variety of building points, such as sensors, setpoints, and commands from HVAC equipment and systems installed in or embedded in the building. The metadata used for modelling purposes is obtained from the Data Clearing House (DCH)~\cite{goldsworthy2022cloud}, a data platform for managing building data using semantic models of buildings. The total number of entities in the original Brick model is 1133. We map and extend this model to RECore, Haystack, and DB.

We also evolve these data points to ensure each category contains various types of instances (individuals). For the equivalent classes within one ontology (e.g., \textit{Break\_Room} and \textit{Breakroom} in Brick), we only create one individual instance. For the synonym classes across building ontologies (e.g., \textit{Floor} in Brick and \textit{Level} in RECore), we have a mapping process and create the same named individual in each ontology. Where these individuals cannot find an exact subclass, we categorise them into the closest superclass (e.g., the zone temperature sensor is mapped to \textit{sensor} in Haystack).

\subsubsection{Building Spaces}

The concept of building spaces is twofold: spaces surrounding the building and spaces within the building. For spaces surrounding the building, we capture the geographical information (e.g., region, site) and outside facilities (e.g., onsite generation, storage, parking). The spaces within the building include physical spaces (e.g., floor and room) and logical spaces (e.g., HVAC zone). Note that the definition of the room is generalised and also includes common spaces (e.g., hallway, lobby, and lounge). Based on the levels of detail in nature, rooms are categorised into different superclasses and subclasses. For example, Room B.0, 1.12, and G.20 are categorised into Room (unspecified), Office (semi-specified), and Enclosed\_Office (specified) classes, respectively. A room is usually assigned to an HVAC zone. In particular, Room B.16 is a service room and does not belong to any HVAC zone.

\subsubsection{Building Equipment and Systems}

The majority of building equipment consists of HVAC equipment. We consider water coil based air conditioning systems that include air side and water side equipment: air handling units (AHU), boilers, chillers, return and supply air fans, etc. The AHU is a large unit responsible for filtering, heating, and cooling the air supply. The basic components include outside and return dampers, filters, heating and cooling valves and coils, and supply fans. Exhaust fans are installed in rooms that require additional air discharge (e.g., the server room and kitchen), while fresh air fans are used for rooms that require additional air supply (e.g., the conference room and enclosed office).

\begin{figure}[!hbt]
\centering
\includegraphics[width=0.95\linewidth]{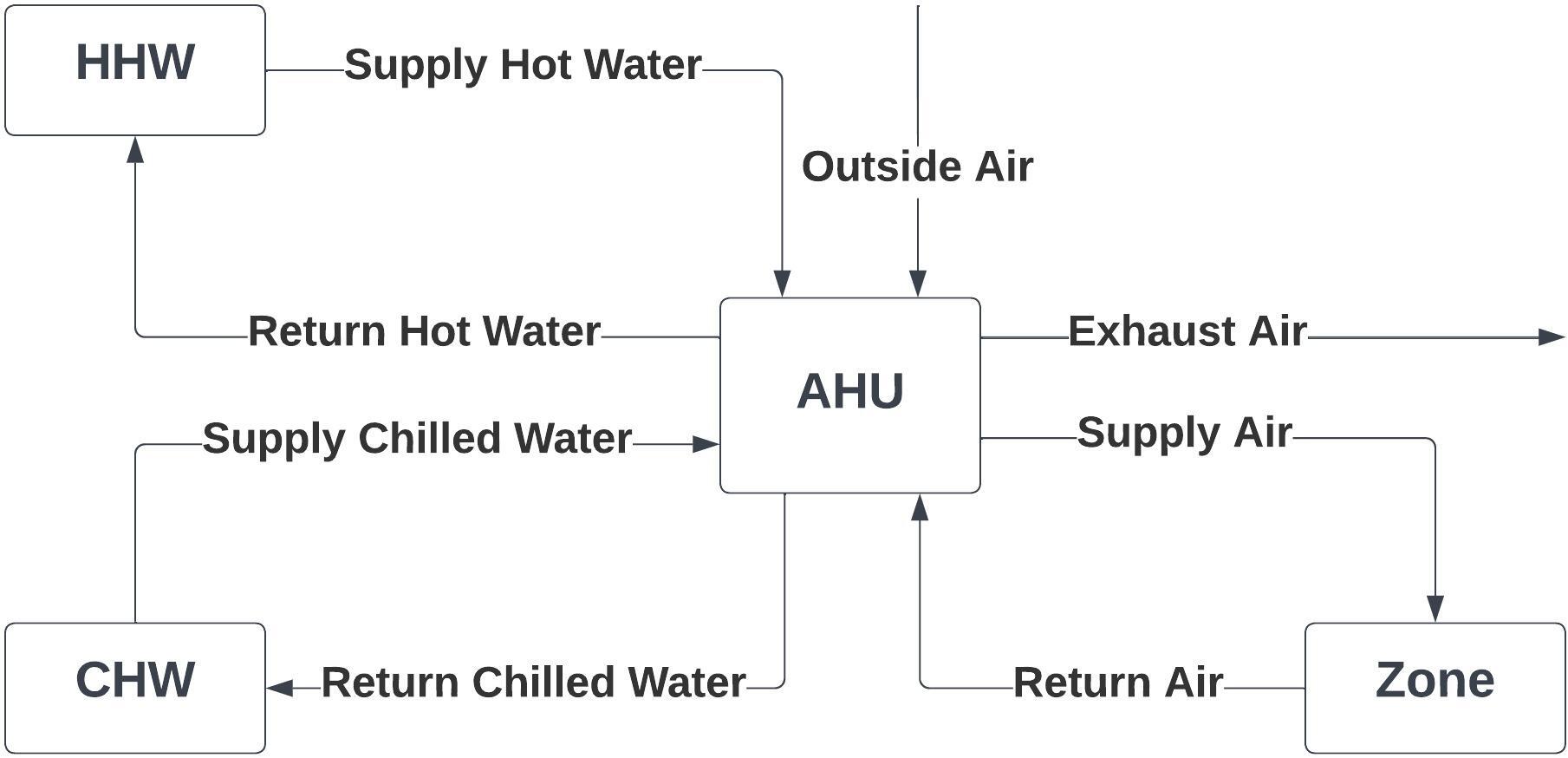}
\caption{The schematics of HVAC systems.}
\label{fig: schematic}
\end{figure}

Fig.~\ref{fig: schematic} illustrates the schematic of HVAC systems. All the data from these systems is collected through a centralised BMS. AHU systems usually include dampers, filters, heating and cooling coils, and supply fans. The heating hot water (HHW) and chilled water (CHW) systems include boilers, chillers, pumps, heating and cooling valves and coils. Note that heating and cooling coils are the shared components across these systems.

\subsubsection{Building Points and Measurements}

We provide various building points operating for different purposes, including alarm, command, parameter, sensor, setpoint, and status. Most of the alarm points appear in the fire panel, while setpoint and status are related to the systems and sub-systems. Command, parameter, and sensor points can be seen in all kinds of building entities (e.g., building spaces, equipment, and systems). These building points measure a wide range of building information and provide useful data for BMS. The classification of building points is defined disparately in different building ontologies. Based on the functions and outputs, we map the same individuals into the corresponding classes. We also provide a variety of measurements at the building points, including temperature, humidity, pressure, etc.

\section{Results}
\label{sec: results}

The results are twofold: (1) the TBox comparison and (2) the ABox comparison. The data models used in this study are available at the following GitHub repository: \url{https://github.com/qzc438/Building-Ontology-Comparison-and-Evaluation} (access will be made available on request)

\subsection{The TBox Comparison of Building Ontologies}
\label{sec: tbox}

\begin{table*}[!hbt]
\centering
\caption{The evaluation criteria of the metrics (retrieved from~\cite{duque2011oquare}).}
\label{tab: evaluation criteria of metrics}
\begin{tabular}{|l|lllll|}
\hline
\multirow{2}{*}{Metrics} & \multicolumn{5}{l|}{Scores} \\ \cline{2-6} 
& \multicolumn{1}{l|}{1} & \multicolumn{1}{l|}{2} & \multicolumn{1}{l|}{3} & \multicolumn{1}{l|}{4} & \multicolumn{1}{l|}{5} \\ \hline
M1-Lack of Cohesion in Methods (LCOMOnto)   & \multicolumn{1}{l|}{\textgreater 8}  & \multicolumn{1}{l|}{(6-8{]}}     & \multicolumn{1}{l|}{(4,6{]}}     & \multicolumn{1}{l|}{(2, 4{]}}    & \textless{}=2     \\ \hline
M2-Weighted Method Count (WMCOnto)          & \multicolumn{1}{l|}{\textgreater 15} & \multicolumn{1}{l|}{(11,15{]}}   & \multicolumn{1}{l|}{(8,11{]}}    & \multicolumn{1}{l|}{(5, 8{]}}    & \textless{}=5     \\ \hline
M3-Depth of subsumption hierarchy (DITOnto) & \multicolumn{1}{l|}{\textgreater 8}  & \multicolumn{1}{l|}{(6-8{]}}     & \multicolumn{1}{l|}{(4,6{]}}     & \multicolumn{1}{l|}{(2, 4{]}}    & {[}1,2{]}         \\ \hline
M4-Number of Ancestor Classes (NACOnto)     & \multicolumn{1}{l|}{\textgreater 8}  & \multicolumn{1}{l|}{(6-8{]}}     & \multicolumn{1}{l|}{(4,6{]}}     & \multicolumn{1}{l|}{(2, 4{]}}    & {[}1,2{]}         \\ \hline
M5-Number of Children (NOCOnto)             & \multicolumn{1}{l|}{\textgreater 12} & \multicolumn{1}{l|}{(8-12{]}}    & \multicolumn{1}{l|}{(6,8{]}}     & \multicolumn{1}{l|}{(3,6{]}}     & {[}1,3{]}         \\ \hline
M6-Coupling between Objects (CBOOnto)       & \multicolumn{1}{l|}{\textgreater 8}  & \multicolumn{1}{l|}{(6-8{]}}     & \multicolumn{1}{l|}{(4,6{]}}     & \multicolumn{1}{l|}{(2, 4{]}}    & {[}1,2{]}         \\ \hline
M7-Response for a Class (RFCOnto)           & \multicolumn{1}{l|}{\textgreater 12} & \multicolumn{1}{l|}{(8-12{]}}    & \multicolumn{1}{l|}{(6-8{]}}     & \multicolumn{1}{l|}{(3-6{]}}     & {[}1-3{]}         \\ \hline
 M8-Number of properties (NOMOnto)          & \multicolumn{1}{l|}{\textgreater 8}  & \multicolumn{1}{l|}{(6-8{]}}     & \multicolumn{1}{l|}{(4,6{]}}     & \multicolumn{1}{l|}{(2, 4{]}}    & \textless{}=2     \\ \hline
M9-Tangledness (TMOnto)                     & \multicolumn{1}{l|}{\textgreater 8}  & \multicolumn{1}{l|}{(6-8{]}}     & \multicolumn{1}{l|}{(4,6{]}}     & \multicolumn{1}{l|}{(2, 4{]}}    & (0,2{]}           \\ \hline
M10-Relationship Richness (RROnto)          & \multicolumn{1}{l|}{{[}0,20{]}\%}    & \multicolumn{1}{l|}{(20-40{]}\%} & \multicolumn{1}{l|}{(40-60{]}\%} & \multicolumn{1}{l|}{(60-80{]}\%} & \textgreater 80\% \\ \hline
M11-Properties Richness (PROnto)            & \multicolumn{1}{l|}{{[}0,20{]}\%}    & \multicolumn{1}{l|}{(20-40{]}\%} & \multicolumn{1}{l|}{(40-60{]}\%} & \multicolumn{1}{l|}{(60-80{]}\%} & \textgreater 80\% \\ \hline
M12-Attribute Richness (AROnto)             & \multicolumn{1}{l|}{{[}0,20{]}\%}    & \multicolumn{1}{l|}{(20-40{]}\%} & \multicolumn{1}{l|}{(40-60{]}\%} & \multicolumn{1}{l|}{(60-80{]}\%} & \textgreater 80\% \\ \hline
M13-Annotation Richness (ANOnto)            & \multicolumn{1}{l|}{{[}0,20{]}\%}    & \multicolumn{1}{l|}{(20-40{]}\%} & \multicolumn{1}{l|}{(40-60{]}\%} & \multicolumn{1}{l|}{(60-80{]}\%} & \textgreater 80\% \\ \hline
M14-Relationships per class (INROnto)       & \multicolumn{1}{l|}{{[}0,20{]}\%}    & \multicolumn{1}{l|}{(20-40{]}\%} & \multicolumn{1}{l|}{(40-60{]}\%} & \multicolumn{1}{l|}{(60-80{]}\%} & \textgreater 80\% \\ \hline
\end{tabular}
\end{table*}

\begin{table*}[!hbt]
\centering
\caption{The associations between characteristics, sub-characteristics, and metrics (evolved from~\cite{tibaut2022framework}).}
\label{tab: associations}
\begin{tabular}{|l|l|l|}
\hline
Characteristics                             & Sub-characteristics                       & Metrics Included           \\ \hline
\multirow{4}{*}{Structural (S)}             & Formal Relations Support (S-FRS)          & M10                        \\ \cline{2-3} 
                                            & Cohesion (S-C)                            & M1                         \\ \cline{2-3} 
                                            & Tangledness (S-T)                         & M9                         \\ \cline{2-3} 
                                            & Redundancy (S-R)                          & M13                        \\ \hline
\multirow{9}{*}{Functional Adequacy (F)}    & Controlled Vocabulary (F-CV)              & M13                        \\ \cline{2-3} 
                                            & Schema and Value Reconciliation (F-SVR)   & M10, M12, M13              \\ \cline{2-3} 
                                            & Consistent Search and Query (F-CSQ)       & M10, M12, M14              \\ \cline{2-3} 
                                            & Knowledge Acquisition (F-KA)              & M13, M10, M8               \\ \cline{2-3} 
                                            & Similarity (F-S)                          & M10, M12                   \\ \cline{2-3} 
                                            & Indexing and Linking (F-IL)               & M10, M12, M14              \\ \cline{2-3} 
                                            & Results Representation (F-RR)             & M12                        \\ \cline{2-3} 
                                            & Guidance and Decision Trees (F-GDT)       & M14, M12                   \\ \cline{2-3} 
                                            & Knowledge Use (F-KU)                      & M13, M12, M14, M8, M1, M4  \\ \hline
\multirow{6}{*}{Maintainability (M)}        & Modularity (M-M)                          & M2, M6                     \\ \cline{2-3} 
                                            & Reusability (M-R)                         & M2, M3, M5, M7, M8, M6     \\ \cline{2-3} 
                                            & Analysability (M-A)                       & M2, M3, M1, M7, M8, M6     \\ \cline{2-3} 
                                            & Changeability (M-C)                       & M2, M3, M1, M7, M8, M6, M5 \\ \cline{2-3} 
                                            & Modification Stability (M-MS)             & M2, M6, M1, M7, M5         \\ \cline{2-3} 
                                            & Testability (M-T)                         & M2, M3, M1, M7, M8, M6     \\ \hline
                Transferability (T)         & Adaptability (T-A)                        & M2, M3, M7, M6             \\ \hline
\multirow{2}{*}{Reliability (R)}            & Recoverability (R-R)                      & M2, M3, M8, M1             \\ \cline{2-3} 
                                            & Availability (R-A)                        & M1                         \\ \hline
                Compatibility (C)           & Replaceability (C-R)                      & M2, M3, M5, M8             \\ \hline
                Operability (O)             & Learnability (O-L)                        & M2, M1, M7, M8, M6, M5     \\ \hline
\end{tabular}
\end{table*}

\begin{figure*}[!hbt]
\centering
\includegraphics[width=1\linewidth]{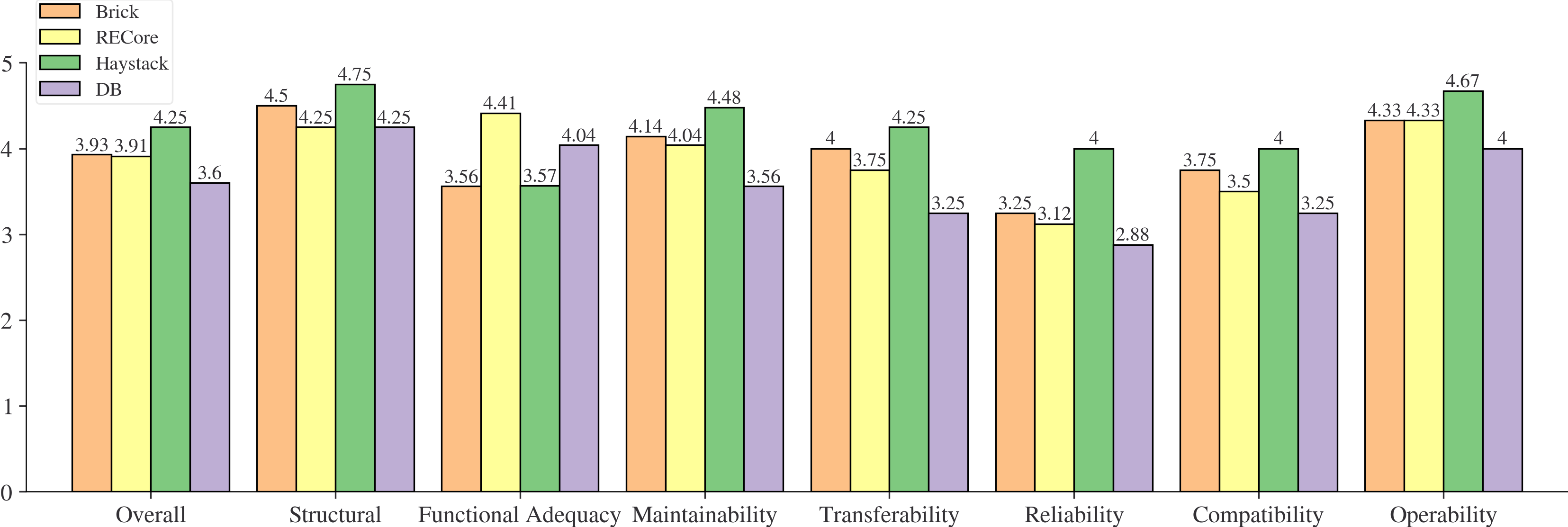}
\caption{The TBox comparison of four building ontologies.}
\label{fig: comparison of four building ontologies}
\end{figure*}

\begin{figure*}[!hbt]
\centering
\includegraphics[width=1\linewidth]{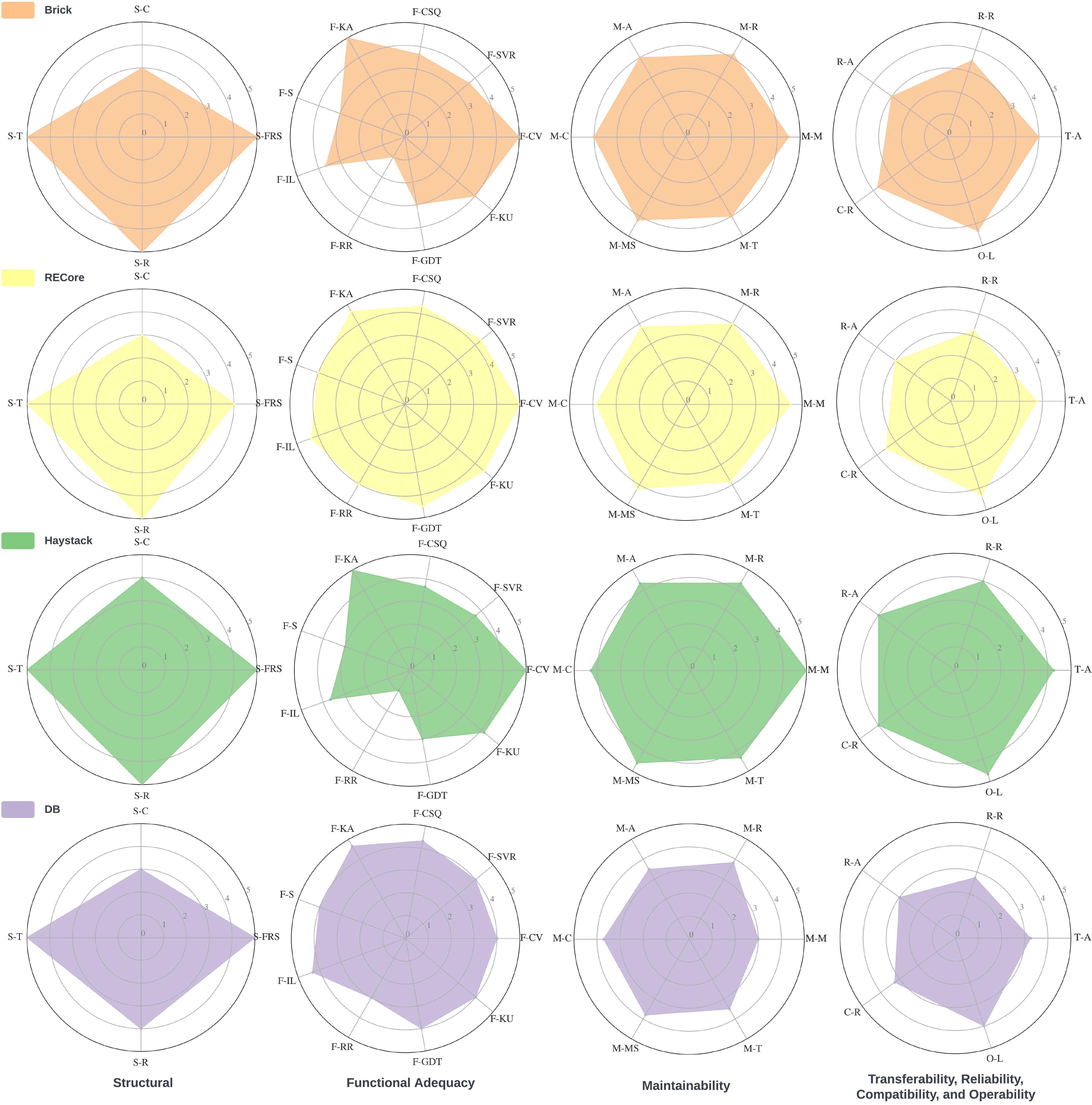}
\caption{The characteristics and sub-characteristics in each building ontology.}
\label{fig: OQuaRE sub-characteristics}
\end{figure*}

\subsubsection{Metrics, Characteristics and Sub-characteristics}

We follow the metrics and evaluation criteria defined in~\cite{duque2011oquare}, while Class Richness (CROnto) is removed and replaced by Relationship Richness (RROnto). CROnto is an ABox evaluation metric that calculates the mean number of instances per class, which is not within the scope of the TBox evaluation. The metrics used in the paper are listed as follows, and their evaluation criteria are shown in Table~\ref{tab: evaluation criteria of metrics}. Note that the scores are scaled from 1 to 5: 1-Not Acceptable, 2-Improvement Required, 3-Minimally Acceptable, 4-Acceptable, and 5-Exceeds Requirements.

\begin{enumerate}[wide, itemsep=0pt, topsep=0pt, labelindent=0pt, label=(\roman*)]
\item M1-Lack of Cohesion in Methods (LCOMOnto): Mean length of the paths from leaf classes to Thing (owl:Thing).
\item M2-Weighted Method Count (WMCOnto): Mean number of the path (edges between classes) from leaf classes to Thing (owl:Thing).
\item M3-Depth of subsumption hierarchy (DITOnto): Length of the largest path from leaf classes to Thing (owl:Thing).
\item M4-Number of Ancestor Classes (NACOnto): Mean number of direct superclasses per leaf class.
\item M5-Number of Children (NOCOnto): Mean number of direct subclasses per leaf class.
\item M6-Coupling between Objects (CBOOnto): Number of direct superclasses of all classes divided by the number of classes not using Thing (owl:Thing) as a direct subclass.
\item M7-Response for a Class (RFCOnto): Number of properties (owl:ObjectProperty and owl:DatatypeProperty) directly accessed from the superclass divided by the number of all classes.
\item M8-Number of properties (NOMOnto): Mean number of properties (owl:ObjectProperty and owl:DatatypeProperty) divided by the number of all classes.
\item M9-Tangledness (TMOnto): Mean number of direct superclasses of the classes with multiple inheritance (i.e., have more than one subclass).
\item M10-Relationship Richness (RROnto): Number of subconcepts (rdfs:subClassOf) defined in the ontology divided by the number of relationships (rdfs:subClassOf) and properties (owl:ObjectProperty and owl:DatatypeProperty).
\item M11-Properties Richness (PROnto): Number of object and data properties defined in the ontology divided by the number of relationships (rdfs:subClassOf) and properties (owl:ObjProperty and owl:DataProperty).
\item M12-Attribute Richness (AROnto): Mean number of property restrictions (owl:Restrictions) per class.
\item M13-Annotation Richness (ANOnto): Mean number of annotations (rdfs:comment and rdfs:label) per class.
\item M14-Relationships per concept: (INROnto): Mean number of relationships (rdfs:subClassOf) per class.
\end{enumerate}

We keep the 5 characteristics from the original paper~\cite{duque2011oquare}: Structural (S), Functional Adequacy (F), Maintainability (M), Reliability (R), and Operability (O). We also add 2 new characteristics, Transferability (T) and Compatibility (C), related to the model transformation and modification process. Some of the sub-characteristics under these 7 characteristics have evolved. Table~\ref{tab: associations} summarises the associations between characteristics, their sub-characteristics, and the metrics involved.

\subsubsection{Axiomatic Design Evaluation}

Fig.~\ref{fig: comparison of four building ontologies} shows the results of comparing four building ontologies. The overall quality scores of the four building ontologies are all above the Minimally Acceptable level (3). Haystack (4.25) has the highest overall score, followed by Brick (3.93), RECore (3.91), and DB (3.60). The scores of the characteristics in each ontology also meet the Minimally Acceptable level (3), except for the score of Reliability (2.88) in DB. We include the detailed scores of characteristics and sub-characteristics of four building ontologies in Fig.~\ref{fig: OQuaRE sub-characteristics} to explore the insights of the axiomatic design.

\begin{enumerate}[wide, itemsep=0pt, topsep=0pt, labelindent=0pt, label=(\roman*)]

\item In Brick, 4 of 7 characteristic scores are equal to or above the Acceptable level (4). They are Structural (4.50), Maintainability (4.14), Transferability (4), and Operability (4.33). However, the characteristic scores of Functional Adequacy (3.56), Reliability (3.25), and Compatibility (3.75) are at the Minimally Acceptable level (3). The low score for Functional Adequacy is due to the sub-characteristics Similarity (F-S) and Results Representation (F-RR) mainly affected by the metric M12-Attribute Richness (AROnto), which means the attributes and axioms defined in each class are relatively insufficient. The low score for Reliability and Compatibility is mainly caused by the metric M3-Depth of subsumption hierarchy (DITOnto). Some classes defined by Brick are relatively nested.

\item In RECore, 4 of 7 characteristic scores are equal to or above the Acceptable level (4). They are Structural (4.25), Functional Adequacy (4.41), Maintainability (4.04), and Operability (4.33). However, the characteristic scores of Transferability (3.75), Reliability (3.12), and Compatibility (3.50) are at the Minimally Acceptable level (3). The low score is due to the sub-characteristics Adaptability (T-A), Recoverability (R-R), Availability (R-A), and Replaceability (C-R) mainly affected by the metrics M1-Lack of Cohesion in Methods (LCOMOnto), M2-Weighted Method Count (WMCOnto), and M3-Depth of subsumption hierarchy (DITOnto). Interestingly, the low scores for LCOMOnto and WMCOnto also indicate the average depth of classes in RECore is even deeper than Brick. However, this does not affect the characteristic of Functional Adequacy in RECore because it is neutralised by the high score of the metric M12-Attribute Richness (AROnto).

\item In Haystack, 6 of 7 characteristic scores are equal to or above the Acceptable level (4). In particular, 3 characteristic scores of Structural (4.75), Maintainability (4.48), and Operability (4.67) reach a higher score and are close to 5. This means the ontology design of Haystack is easy to maintain and operate. Similar to Brick, the low score for Functional Adequacy is affected by the metric M12-Attribute Richness (AROnto), which means the attributes and axioms defined in each class are relatively insufficient.

\item In DB, 3 of 7 characteristic scores are equal to or above the Acceptable level (4). They are Structural (4.25), Functional Adequacy (4.04), and Operability (4). The other 3 characteristic scores of Maintainability (3.56), Transferability (3.25), and Compatibility (3.25) are at the Minimally Acceptable level (3); in particular, the score of Reliability (2.88) is below the Minimally Acceptable level (3). It is caused by the low scores of the metrics M1-Lack of Cohesion in Methods (LCOMOnto), M2-Weighted Method Count (WMCOnto), and M3-Depth of subsumption hierarchy (DITOnto). DB commonly uses the delimiter ``\_'' to separate the class name into several components in line with the original design of the BMS, but it also results in a deeper hierarchy. For example, the equipment class Multi-directional Shade (East and Southeast) (\url{http://www.google.com/DB/0.0.1/hvac#Sdc_ext_tlt_east_southeast}) has a depth of 7 (EntityType-$>$Equipment-$>$Sdc-$>$Sdc\_ext-$>$Sdc\_ext\_tlt-$>$Sdc\_ext\_tlt\_east-$>$Sdc\_ext\_tlt\_east\_southeast).

\end{enumerate}

\subsubsection{Summary}

The TBox evaluation shows Haystack and Brick perform better than RECore and DB on the assessed metrics. This is because Haystack and Brick tend to be wide class hierarchies, while RECore and DB tend to be deep class hierarchies. Ontology is defined as ``a formal specification of a conceptualisation". It abstracts the concepts and their relations to merge, share, and reuse presented knowledge or deduce facts that do not exist in the knowledge base~\cite{gruninger1995methodology}. In this setting, the design of the ontology should not only aim for high coverage of the knowledge (i.e., completeness and expressiveness) but also target efficiency by minimising the classes, properties, and axioms used, and avoiding redundant definitions and an excessively deep class hierarchy, so that it can be easily used for querying and reasoning. Ideally, an ontology that scores well on TBox metrics is more compact. This is useful for applications where querying and inferencing are simple, or for IoT devices with limited data transfer capacity. We also find the following implications:

\begin{enumerate}[wide, itemsep=0pt, topsep=0pt, labelindent=0pt, label=(\roman*)]
\item Although the ontological representation of the building data helps with the Structural (S) and Operability (O), all four building ontologies observe a relatively deep and nested class hierarchy. This affects the Functional Adequacy (F) of the ontology, as well as the capabilities of Maintainability (M), Transferability (T), Reliability (R), and Compatibility (C).
\item The use of attributes and properties (usually observed in ontologies with deep class hierarchies) could help with the Functional Adequacy (F), but it also has negative impact on the Maintainability (M), Transferability (T), Reliability (R), and Compatibility (C). The complexity of the ontology class definitions makes it hard to maintain and exchange data between different building ontologies.
\item The use of subclasses and annotations (usually observed in ontologies with wide class hierarchies) could help with the Maintainability (M), Transferability (T), Reliability (R), and Compatibility (C). Such a design makes it easier to discover the nested classes, while querying and inferring the facts in the ontology may have some issues with the Functional Adequacy (F). The fact that simple ontology design decisions have mutually contradictory effects on metrics demonstrates the difficulty of balancing competing needs for creators and users of models.
\end{enumerate}

\subsection{The ABox Evaluation of Building Ontologies}
\label{sec: abox}

\subsubsection{Concept Evaluation}

\paragraph{A. Building Spaces}

\begin{figure*}[!hbt]
\centering
\includegraphics[width=1\linewidth]{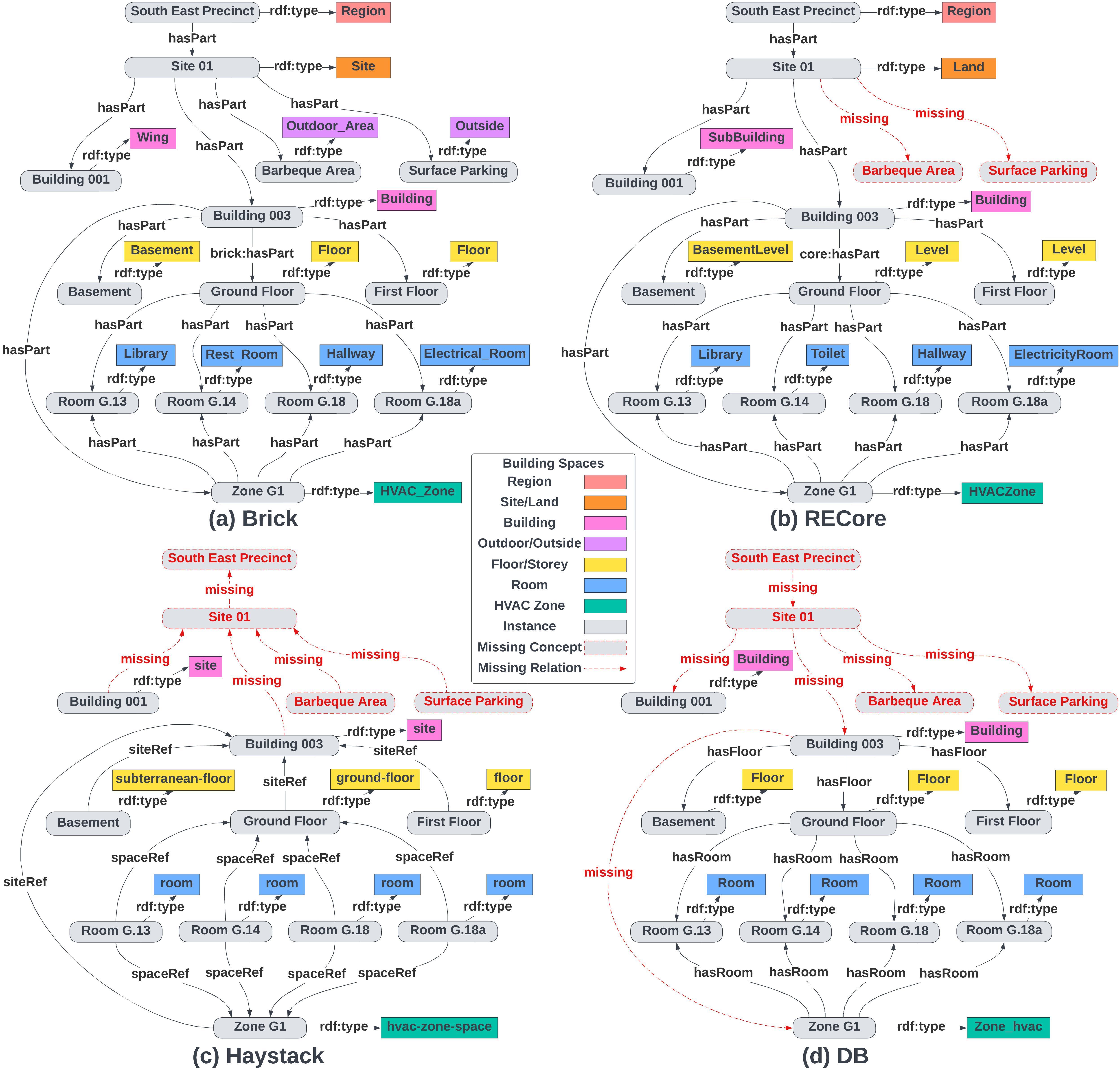}
\caption{Comparison of four building ontologies representing building spaces.}
\label{fig: building spaces}
\end{figure*}

A building space is characterised as a contiguous part of the building with a 3D spatial extent. It can be bound physically or conceptually. An example of a building space is described as follows: \textit{``Building 003, located on the South East Precinct Site 01, is an office building with a wing Building 001. Outside the building, there is a barbeque area and surface parking. The building has three floors: Basement, Ground Floor, and First Floor. The library (Room G.13) is on the Ground Floor. It is located to the south of the restroom (Room G.14), east of the hallway (Room G.18), and opposite to the electrical room (Room G.18a). These four rooms belong to HVAC Zone G1.''} Fig.~\ref{fig: building spaces} shows the model for the building spaces of our use case in each of the four ontologies.

\begin{enumerate}[wide, itemsep=0pt, topsep=0pt, labelindent=0pt, label=(\roman*)]
\item There are different terms referring to the building spaces in the four ontologies investigated. In Brick and RECore, building spaces are categorised into \textit{Location} and \textit{Space}, respectively. The spatial information of Haystack is classified into \textit{site} and \textit{space}. DB separates the definition of building spaces into two classes: \textit{PhysicalLocation} (physical spaces) and \textit{Zone\_hvac} (logic spaces).
\item All of the four building ontologies are able to capture the basic concepts of building spaces, such as buildings, floors, rooms, and HVAC zones. Nevertheless, they have different levels of expressiveness in representing the building-related artifacts. Brick could capture the concepts of the region, site (land) and outdoor area (outside), while the concept of the outdoor area (outside) is missing in RECore. Haystack and DB assume one building per site (although Haystack has a site class, it is defined as one building with its own unique street address). They cannot describe the spatial information related to region, site (land), and outdoor area (outside).
\item These concepts are linked by part-whole relationships. The part-whole relationship can be made and maintained once and reused plenty of times. For example, the floor is a part of the building, the room is a part of the floor, and the room is a part of the HVAC zone, etc. This idea was introduced in BOT~\cite{rasmussen2017proposing}. In Brick and RECore, the transitive property \textit{hasPart} is used. In Haystack, two inverse properties \textit{siteRef} and \textit{spaceRef} are defined. Unlike other ontologies, DB has one property \textit{hasPhysicalLocation}, and two sub-properties \textit{hasFloor} and \textit{hasRoom}. Note that no appropriate properties can be found to describe the relationship between logic zones in DB.

\begin{figure*}[!hbt]
\centering
\includegraphics[width=1\linewidth]{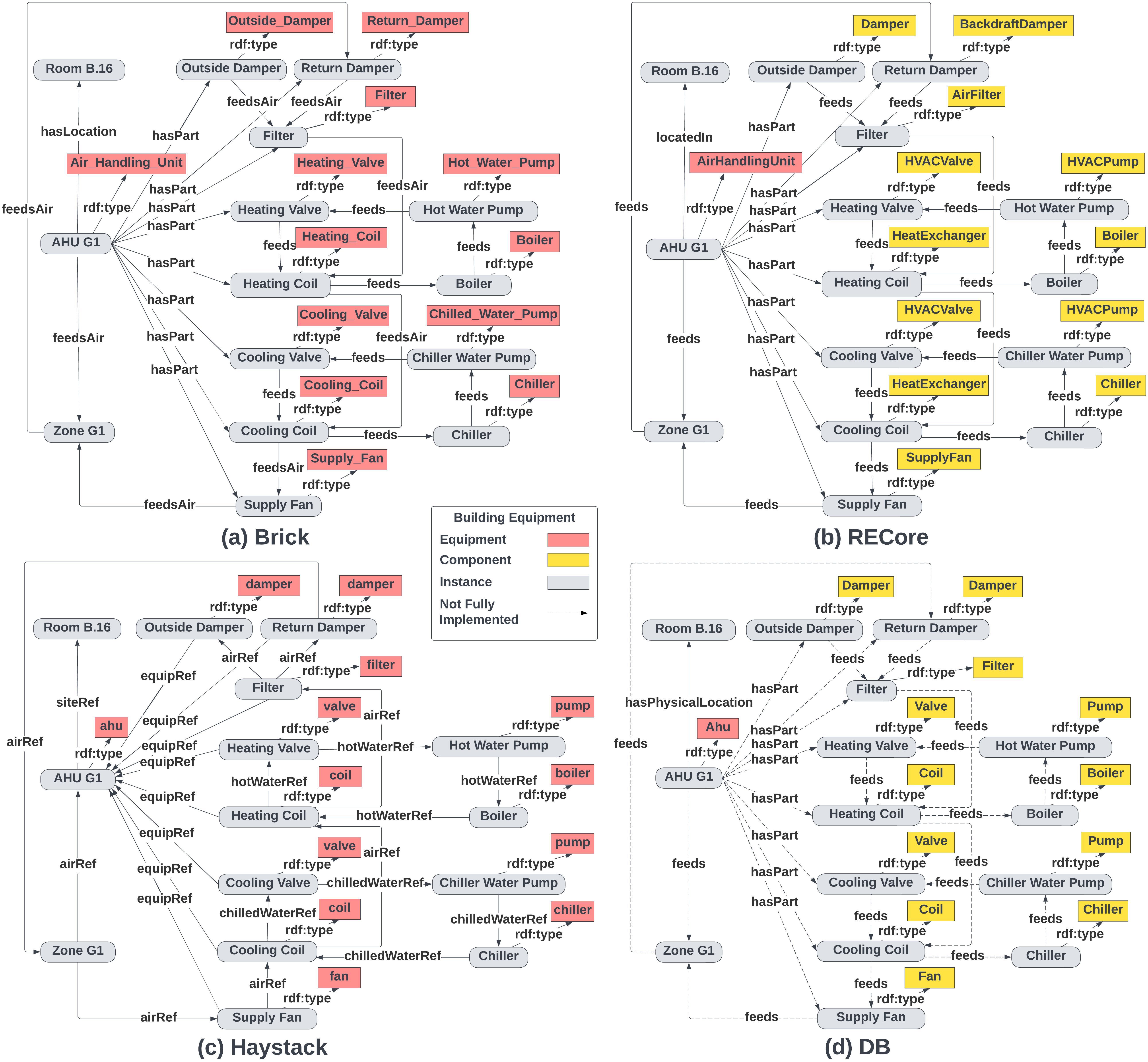}
\caption{Comparison of four building ontologies representing building equipment and systems (a).}
\label{fig: equipment and system (a)}
\end{figure*}

\begin{figure*}[htbp]
\centering
\adjustbox{max width=\linewidth,valign=m}{\includegraphics{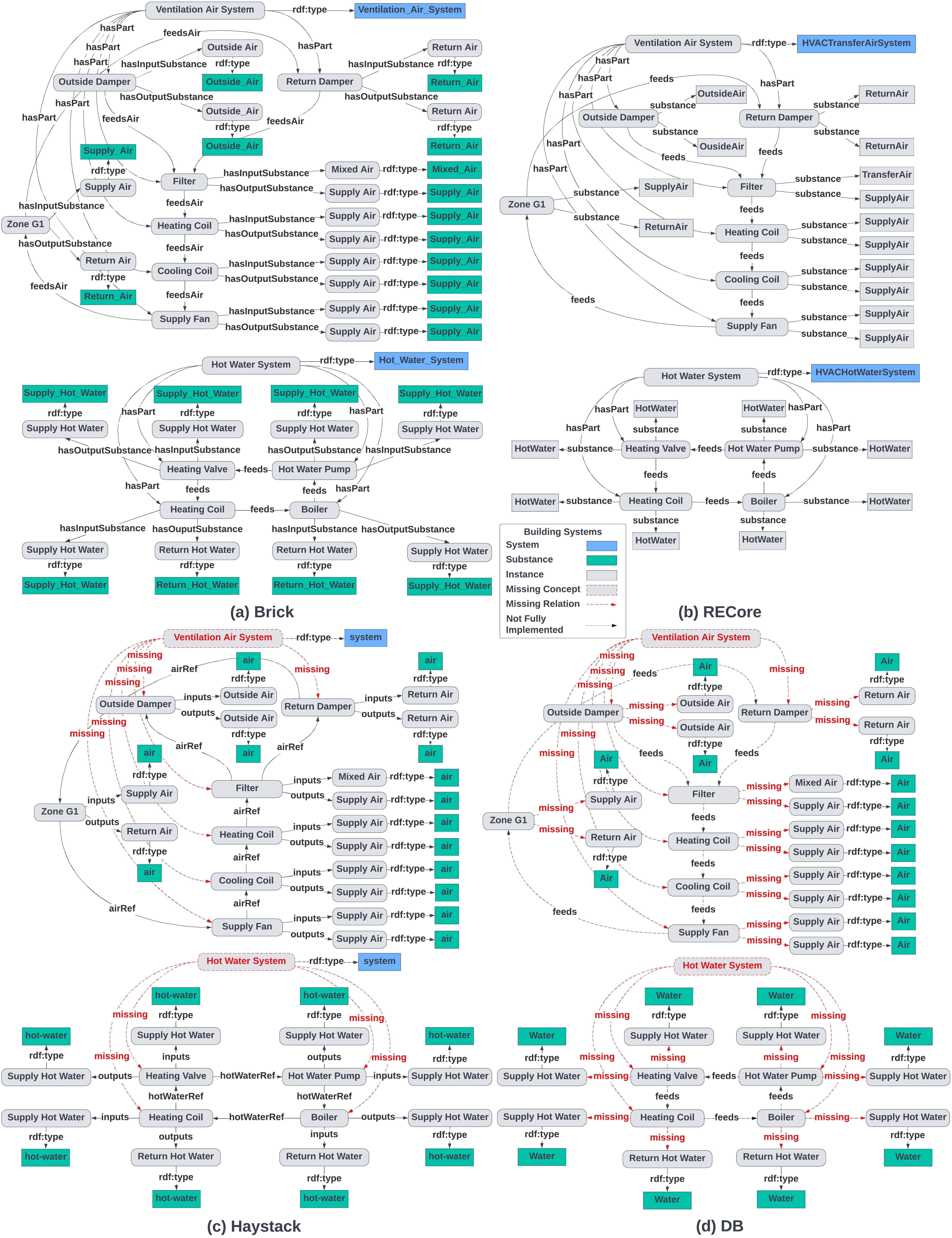}}
\vspace{-0.4cm}
\caption{Comparison of four building ontologies representing building equipment and systems (b).}
\label{fig: equipment and system (b)}
\end{figure*}

\item The functions of a room are expressed at different abstraction levels in the four building ontologies. Brick and RECore specify the room functions by defining new classes and subclasses for each function. In contrast, Haystack and DB only provide the general class \textit{Room}. Categories are not used in any of the ontologies.
\item All of the four building ontologies have limited expressiveness in representing the geometry of the building space (e.g., sizes, shapes, positions, angles, and dimensions) and topological connections between the building spaces (e.g., contain, adjacent, and intersect). Only RECore provides some limited functions to represent the geometry using the class \textit{geo:Geometry} and property \textit{geo:hasSerialization} from GeoSPARQL~\cite{battle2011geosparql}. It is possible to link building ontologies to BOT~\cite{rasmussen2021bot} and GeoSPARQL~\cite{battle2011geosparql}.
\end{enumerate}

\paragraph{B. Building Equipment and Systems}

Building equipment refers to devices that are (semi-) permanently mounted or installed in a building space, and measure, transport, or transform either materials or energy across the whole or part of the building. A system is a collection of equipment. An example of equipment and its related systems is described as follows: \textit{``AHU G1 is located in the service room (B.16) and feeds air to HVAC Zone G1. The basic components of the AHU G1 are the outside damper, return damper, filter, heating valve, cooling valve, heating coil, cooling coil, and supply fan. The AHU comprises the air system and is connected to the water system. The air system contains outside damper, return damper, filter, heating coil, cooling coil, and supply fan. In a ventilation mode, outside air and return air are passed into the ventilation air system through the outside dumper and return dumper. Then, the mixed air goes through the filter and turns into the supply air. After that, air goes through the heating coil and cooling coil (to heat or cool the air if necessary), and finally, the supply fan feeds the supply air to the HVAC Zone G1. The hot water system comprises boiler, hot water pump, heating valve, and heating coil. The supply of hot water comes from the boiler via the hot water pump. It flows through the valve and coil. The valve also modulates the flow depending on the required load (the description of the chilled water system is omitted, similar to the hot water system).''} Fig.~\ref{fig: equipment and system (a)} and Fig.~\ref{fig: equipment and system (b)} show the model for the building equipment and systems of our use case in each of the four ontologies.

\begin{enumerate}[wide, itemsep=0pt, topsep=0pt, labelindent=0pt, label=(\roman*)]

\item The classes used to describe the equipment in the four building ontologies are effectively equivalent. Some domain ontologies use the synonym \textit{Device}, but these four building ontologies all use the term \textit{Equipment}. Only Brick and RECore have the definition of the \textit{System}. Brick also defines \textit{Loop} to describe a collection of connected equipment, such as the air loop and water loop. This class is useful when it comes to primary and secondary loops in air and water systems (an example is shown in Fig.~\ref{fig: system and loop}). Collection classes are not fully implemented in Haystack and DB, including systems and loops.

\begin{figure}[!hbt]
\centering
\includegraphics[width=1\linewidth]{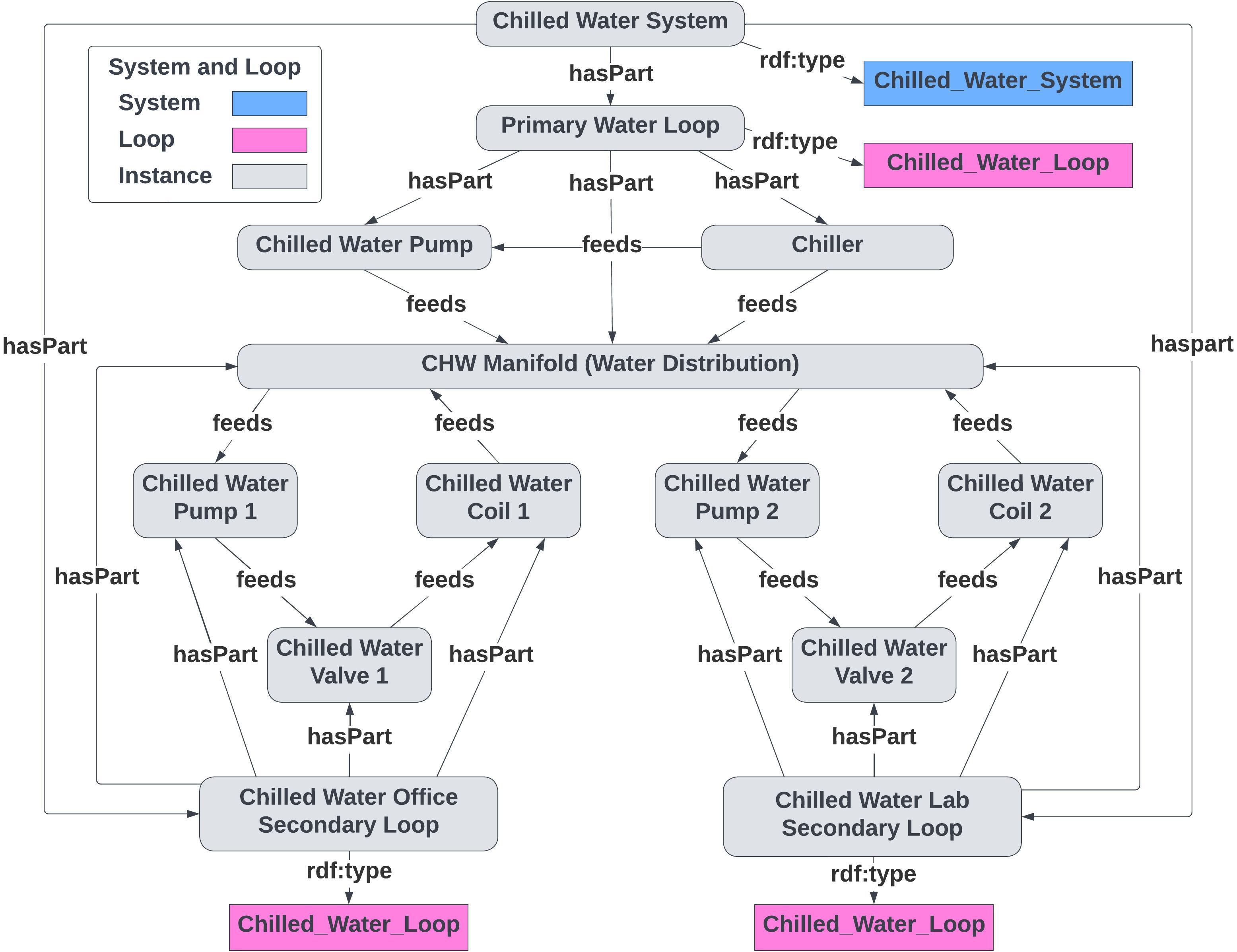}
\caption{An example of system and loop in Brick.}
\label{fig: system and loop}
\end{figure}

\item Brick and Haystack do not have a classification for equipment and sub-equipment. RECore and DB use the \textit{Component} class to describe the sub-equipment, a part of the equipment. In this setting, equipment is mechanical or electrical devices that are functionally independent, while sub-equipment (i.e., component) is manufactured objects that are functionally dependent. For instance, Fan is equipment, but HVAC Fan is a component because it usually works with other building parts.

\item Equipment and system are linked by part-whole and feeds relationships. Some of the documented connection properties in DB are not fully implemented in the ontology at the time of writing. The selection of a relationship property depends on the type of connection between two objects. A logical connection uses the \textit{part-whole} relationship, while a physical connection uses the \textit{feeds} relationship. A physical connection is tangible, where one thing connects to another through piping, wiring, or other similar pathways. Physical connections are built to transmit substances. For example, the connection between the filter and the heating or cooling coil within an AHU is a physical connection. They are connected through a duct, and the air is transmitted between these two objects. The logical connection is intangible and represents an abstract concept or topology of the network. For instance, the connections between the AHU and filter, or between the AHU and heating or cooling coil, are logical connections because the filter and heating or cooling coil are parts of the AHU. Uniquely, Haystack uses reference objects to classify physical and logical connections. \textit{airRef}, \textit{hotWaterRef}, and \textit{chilledWaterRef} are physical connections, whereas \textit{equipmentRef} is used for logical connections.

\item The four building ontologies provide different levels of detail to describe feeds relationships. Rather than using plain \textit{feeds}, the predicates used in Haystack are \textit{airRef}, \textit{chilledWaterRef} and \textit{hotWaterRef}. Unlike Haystack, the property \textit{feeds} in Brick only has one sub-property \textit{feedsAir}, while RECore and DB do not have any sub-properties under \textit{feeds}.

\item It is hard for users with limited engineering knowledge to determine whether a connection is physical or logical. We found an easy way to distinguish the types of connections by using a discrete definition of equipment and component. Connections within the category (equipment and equipment, component and component) are physical (i.e., use \textit{feeds}, \textit{airRef}, \textit{hotWaterRef}, and \textit{chilledWaterRef}), while the connections across dissimilar categories (equipment and component) are logical (i.e., use \textit{hasPart} and \textit{equipmentRef}) in common.

\item All four building ontologies agree that equipment can simultaneously have two linked properties to the building space. The reason is that equipment can be located in one place but feeds into another place or HVAC zone. In our case, AHU G1 is located in the service room (B.116) but feeds the air to Zone G1, an HVAC zone on the Ground Floor.

\item The four building ontologies use different approaches to list substances. Brick and Haystack define new classes and subclasses. RECore uses enumerated individuals, while DB only provides general classes such as \textit{Air} and \textit{Water}. QUDT~\cite{hodgson2011qudt} has a finite list of qualities and could provide a way to ensure the completeness and expressiveness of the substances.

\item Regarding the properties used to link equipment and substances, the four building ontologies also provide different levels of detail. Brick and Haystack have a classification of input and output substances by using the properties \textit{hasInputSubstance} (\textit{input}) and \textit{hasOutputSubstance} (\textit{output}). RECore only has a general \textit{Substance} class. For DB, no appropriate input and output properties can be found.

\end{enumerate}

\begin{figure*}[!hbt]
\centering
\includegraphics[width=1\linewidth]{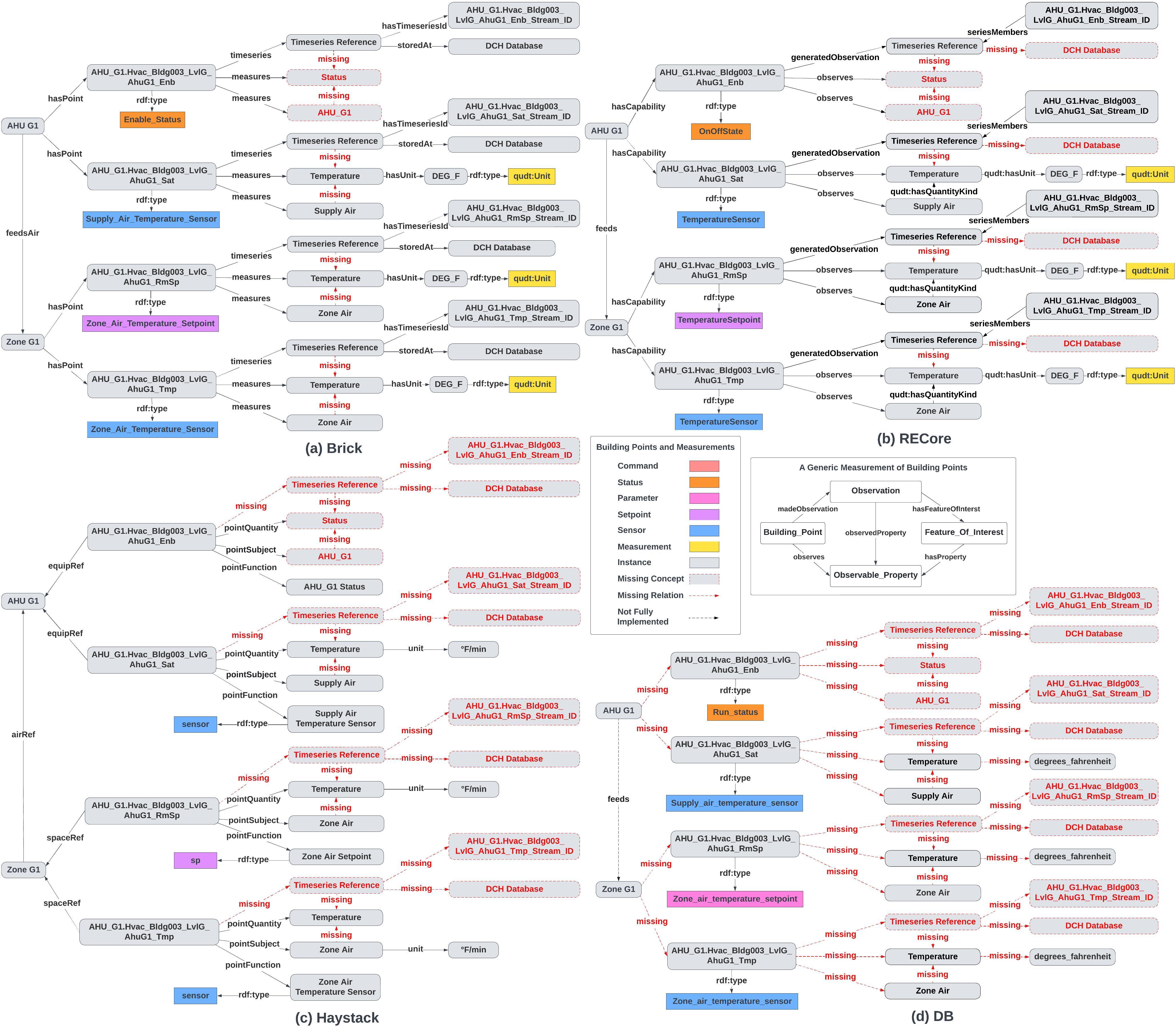}
\caption{Comparison of four building ontologies representing building points and measurements.}
\label{fig: building points and measurements}
\end{figure*}

\paragraph{C. Building Points and Measurements}

Building points refer to sensors or actuators embedded in a building entity (i.e., a building space, equipment, or system). They can be extended to setpoints, states, commands, or parameters of an entity that produce or ingest data. Building points are the points of measurable telemetry and actuation, and measurements usually include the subject, the quantity of the subject, and the unit of measurement. An example of building points and measurements of the air temperature is described as follows: \textit{``AHU G1 feeds air to HVAC Zone G1. To have better control of the air temperature, there are four types of IoT data stored in the DCH database for review and analysis. The system enables status to track the current state of the AHU G1, i.e., whether the AHU G1 is on or off. The supply air temperature sensor is embedded in the AHU G1 and measures the supply air temperature. There are two points located in HVAC Zone G1. The zone air temperature setpoint indicates the desirable setting of the air temperature within the zone, whereas the zone air temperature sensor records the real-time temperature of the zone air.''} Fig.~\ref{fig: building points and measurements} shows the model for the different building points and measurements of our use case in each of the four ontologies.

\begin{enumerate}[wide, itemsep=0pt, topsep=0pt, labelindent=0pt, label=(\roman*)]

\item Among the four building ontologies, there are conflicts in the definition of building points. Brick and Haystack use the \textit{Point} class and the \textit{hasPoint} property. They define building points as measurable data points embedded in building spaces, equipment, and systems that produce and ingest data. RECore, however, uses a \textit{Capability} class and the \textit{hasCapability} property. The reason is that building points provide the capability for an entity to produce or ingest data. Depending on the functions and behaviours, building points can be specialised into different categories. DB defines building points as a field belonging to an entity.

\item There are also dissimilarities in defining the subclasses under \textit{Point} or \textit{Capability}. DB defines each point as a field, a self-descriptive object that consists of a string of subfields, for example, \textit{zone\_air\_temperature\_sensor}. Other ontologies categorise points by their functions. Haystack only has three types of building points: \textit{cmd} (command), \textit{sensor}, and \textit{sp} (setpoint). In Brick and RECore, point types also include the system state (status) and system parameters.

\item In our case, the time-series data is stored in the DCH database. Only Brick provides such a function to store the time series ID in the ontologies and allow the full dataset to be stored independently outside the ontology. This not only reduces the size of the ontology but also enables better management of the time-series data in a proper time-series database. RECore, Haystack, and DB do not have the definition of \textit{Database} and only provide one option to store time-series data directly on the knowledge graph itself.

\item A generic measurement of building points includes \textit{Building\_Point}, \textit{Observation}, \textit{Observable\_Property}, and \textit{Feature\_Of\_Interest}, as described in SSN~\cite{compton2012ssn} and SOSA~\cite{haller2019modular}. Their relationships are shown in the middle of Fig.~\ref{fig: building points and measurements}. As an example, consider measuring zone air temperature. The building points are the zone temperature sensor, the observation is time-series data, the feature of interest is the zone air, and the observable property is the temperature. None of the four building ontologies could capture all the basic components of the measurement. All of them are missing or partially missing the links between \textit{Observation} and \textit{Observable\_Property} (i.e., \textit{observedProperty}), and between \textit{Feature\_Of\_Interest} and \textit{Observable\_Property} (i.e., \textit{hasProperty}). Other properties can find equivalent properties in the four building ontologies. In Brick, the property \textit{measures} is roughly equivalent to \textit{observes} and \textit{hasFeatureOfInterest}. Equivalent properties found in RECore is \textit{observes}. Besides the equivalent properties \textit{pointQuantity} and \textit{pointSubject}, Haystack also uses a property \textit{pointFunction} to specify the point functions.

\item All the equivalent classes of \textit{Observable\_Property} and \textit{Feature\_Of\_Interest} in the four building ontologies are bound to quantities and substances only. As a result, they have limitations in representing the system status (state) or setpoint. In RECore, the domain of the property \textit{observes} is the \textit{Sensor} class only. There are no appropriate properties that can be used for other types of points. In DB, this part is still under development, and related properties are missing.

\item The four building ontologies use two different approaches for the unit of measurement. Brick and RECore reuse the QUDT~\cite{hodgson2011qudt} \textit{Unit} class and its subclasses, while Haystack and DB use the enumerated individuals and list all the possible units within the ontology.

\end{enumerate}

\subsubsection{Use Case Evaluation: Energy Audits}
\label{sec: evaluation}

\begin{table*}[!hbt]
\caption{Energy Audits - The examples of competency questions answered by four building ontologies.} 
\label{tab: energy audits competency questions}
\begin{adjustbox}{width=1\linewidth,center}
\begin{tabular}{|c|l|cccc|}                            \hline
Concepts                                & Examples of Competency Questions  & B   & R   & H   & D       \\ \hline
\multirow{5}{*}{\makecell{Building Spaces}}        
                                        & \#01: What is the primary use of the building? If multiple uses, what is the fraction of floor area for each use?  
                                        & \halfcirc  & \halfcirc & \halfcirc  & \emptycirc  \\
                                        & \#02: What are the HVAC zones and their respective air-conditioned areas in the building?    
                                        & \fullcirc  & \fullcirc & \fullcirc & \halfcirc  \\
                                        & \#03: What are all the rooms and their floors in the building? 
                                        & \fullcirc  & \fullcirc & \fullcirc & \halfcirc  \\
                                        & \#04: What is the function type of each room?   
                                        & \fullcirc  & \fullcirc & \halfcirc  & \halfcirc   \\ 
                                        & \#05: What is the occupancy of each room?                             
                                        & \fullcirc & \fullcirc & \fullcirc  & \emptycirc   \\ \hline
 \multirow{6}{*}{\makecell{Building Equipment \\ \&  Systems}}  
                                        & \#06: What are the pieces of equipment and systems of equipment in the building?                                   
                                        & \fullcirc  & \fullcirc & \halfcirc  & \halfcirc   \\
                                        & \#07: What are the properties (e.g., type, physical location and functional area) of each piece of HVAC equipment?                                         
                                        & \fullcirc  & \fullcirc & \fullcirc & \halfcirc  \\
                                        & \#08: What are the components of each piece of equipment and their structural relations?   
                                        & \fullcirc  & \fullcirc & \fullcirc  & \emptycirc  \\
                                        & \#09: What is the energy use broken down by resource using the energy?               
                                        & \fullcirc  & \halfcirc & \halfcirc & \emptycirc  \\
                                        & \#10: What is the energy use broken down by end-use?
                                        & \fullcirc  & \fullcirc & \halfcirc & \emptycirc  \\
                                        & \#11: What are the energy-generating pieces of equipment in the building?                         
                                        & \halfcirc  & \fullcirc & \emptycirc & \emptycirc  \\ \hline
\multirow{5}{*}{\makecell{Building Points \\ \& Measurements}} 
                                        & \#12: What are the meters and meter types in the building?                              
                                        & \fullcirc  & \fullcirc & \fullcirc  & \emptycirc   \\
                                        & \#13: What does each meter measure (i.e., substance and quantity)?   
                                        & \halfcirc  & \halfcirc & \fullcirc  & \emptycirc  \\
                                        & \#14: What are the connections between meters and sub-meters?           
                                        & \halfcirc  & \halfcirc & \fullcirc & \emptycirc  \\ 
                                        & \#15: What are the operating parameters of each piece of equipment (e.g., setpoints and schedules)?  & \fullcirc  & \fullcirc & \fullcirc  & \emptycirc   \\
                                        & \#16: What are the time series for each meter?      
                                        & \fullcirc  & \fullcirc & \halfcirc  & \emptycirc   \\ \hline
\multicolumn{6}{c}{B - Brick, R - RECore, H - Haystack, D - DB}               \\
\multicolumn{6}{c}{\fullcirc~- Fully Answered, \halfcirc~- Partially Answered, \emptycirc~- Unable to Answer}
\end{tabular}
\end{adjustbox}
\end{table*}

The use case evaluation is based on competency questions retrieved from application-level scenarios. We apply the four building ontologies to the use case and compare their completeness and expressiveness, i.e., how many competency questions can be answered and how well these questions are answered. Note that we set the answers for these competency questions into three categories, and ``Partially Answered'' is a medium between ``Fully Answered'' and ``Unable to Answer''. It means that only part of the information can be found, or that the answers are not explicit but implicit.

An energy audit is the assessment of total energy usage across electricity, water, and gas in a smart building and its energy-consuming systems~\cite{krarti2020energy}. The use of building ontologies enables auditors to query and retrieve the required sampling data. According to ASHRAE Standard 211~\cite{ASHRAE211}, energy audits can be performed at one of three levels: Level 1 (walk-through audit), Level 2 (detailed energy audit), and Level 3 (investment-grade audit). Each level covers the requirements of the previous level and includes additional information. We perform a Level 2 energy audit.

Table~\ref{tab: energy audits competency questions} summarises the examples of competency questions and how well each building ontology can potentially answer these questions\footnotemark. In this use case, Brick and RECore's full and partial answers add up to 18/20, followed by Haystack (16/20) and DB (2.5/20). The main limitations of the building ontologies lie in the lack of building layout presentation (i.e., geometry and floor area). This also impacts the space function breakdown related to energy use and savings. Moreover, Haystack and DB do not specify the space, equipment, and system type, which makes it hard to determine the systematic composition and restricts their use in energy audits.

\footnotetext{Details of the use case evaluation can be found in~\ref{appendix: competency questions in energy audits}.}

\subsubsection{Summary}

The ABox evaluation shows that Brick and RECore potentially outperform Haystack and DB. From a holistic view, Brick and RECore tend to use an extensible class hierarchy to define specific types of building entities, while Haystack and DB usually only define general classes, properties, and instances. For example, the \textit{Room} class in Brick and RECore contains multiple subclasses to distinguish the room type, such as office room, conference room, and rest room. Conversely, Haystack and DB only define one general class \textit{Room}, and the room function is not detailed. This design pattern is common and can be observed in the descriptions of the building entities among all three main concepts. However, there are some exceptions. Haystack and DB also provide more detailed and specific taxonomies for classifying the sensors. In Haystack, the sensor type can be explicitly inferred by the substance (property \textit{pointSubject}) and quantity (property \textit{pointQuantity}) it measures. DB uses the delimiter ``-'' to separate the sensor’s name into several components, and these components are meaningful and may indicate its function (\textit{supply\_air\_temperature\_sensor}). We also find the following implications:

\begin{enumerate}[wide, itemsep=0pt, topsep=0pt, labelindent=0pt, label=(\roman*)]

\item On a macro level, there are well-agreed design patterns observed in building spaces, equipment, and systems. The four building ontologies use different classes and properties to describe the information related to building spaces, equipment, and systems, but their underlying definitions and axioms are similar.

\item On a micro level, conflicting design patterns appear in the four building ontologies, and they are observed in building points and measurements. The four building ontologies have different classifications of the building points, and use different definitions to illustrate the relationship between building points and building spaces, equipment, and systems (i.e., \textit{hasPoint} and \textit{hasCapability}).

\item Reusing concepts defined in existing ontologies is better for interoperability and usability than creating new ones from scratch. We recommend reusing the concepts defined in BOT~\cite{rasmussen2021bot} and GeoSPARQL~\cite{battle2011geosparql} to represent spatial information, SAREF~\cite{daniele2015created} and SAREF building extension~\cite{poveda2018extending} to represent equipment and systems, SSN~\cite{compton2012ssn} and SOSA~\cite{haller2019modular} to represent sensors and actuators, and QUDT~\cite{hodgson2011qudt} for corresponding measurements. These domain ontologies are usually more complete and expressive than the ones created locally in the building ontologies~\cite{lbd2023}.

\end{enumerate}

\section{Discussion}
\label{sec: discussion}

There is no universal building ontology that always performs best and fits all the task requirements. In other words, the no free lunch (NFL) theorem~\cite{wolpert1997no} also applies to ontology design for smart building applications. Referring to our study, each building ontology has its pros and cons and achieves different levels of abstraction. We believe there is no context-independent reason to favour one building ontology over another. When we claim one building ontology is better than another, it is not because of its inherent features, but because of its better fitness for the task requirements. However, as a complement to NFL, comparing and evaluating building ontologies through qualitative and quantitative analysis can be a good reference for determining which building ontology to use.

\begin{figure}[!hbt]
\centering
\includegraphics[width=0.9\linewidth]{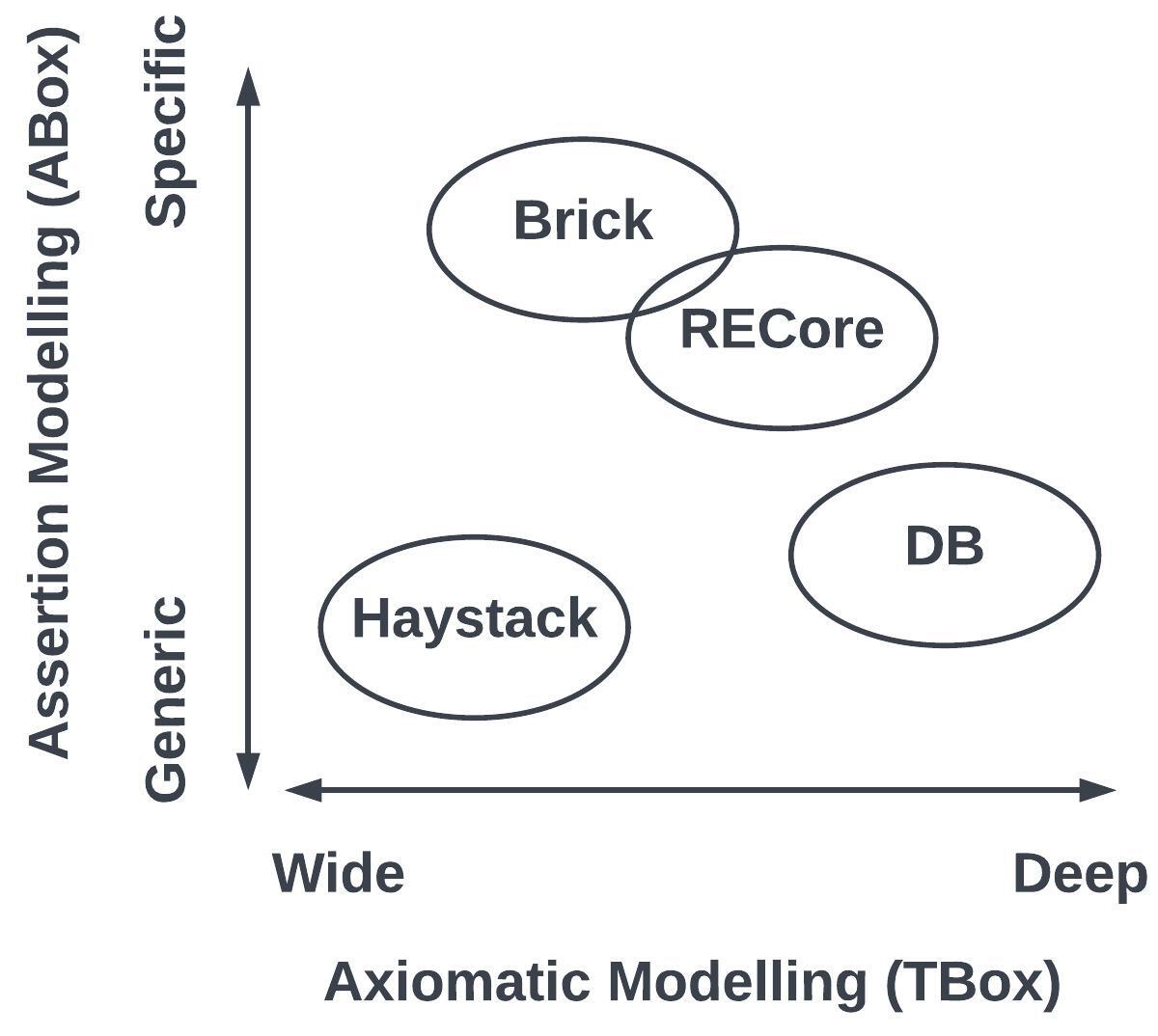}
\caption{The overview of building ontology modelling patterns.}
\label{fig: modelling patterns}
\end{figure}

In this paper, we compare and evaluate four popular building ontologies with respect to both TBox and ABox components. In the TBox evaluation, we apply the OQuaRE Framework to evaluate the axiomatic design of the building ontologies. The results indicate that Haystack and Brick are better than the other two building ontologies with respect to efficiency and compactness. In the ABox evaluation, we apply four popular building ontologies to a specific building case study to conduct a fair comparison. We evaluate three main concepts: building spaces, equipment and systems, and points and measurements. For each concept, we compare and analyse the four building ontologies representing the same use case. We also use a real-world application, energy audits, to evaluate completeness and expressiveness. The results demonstrate that Brick and RECore cover more of the required domain concepts than the other two building ontologies. The conflicting results of the TBox and ABox evaluations reveal that the choice of the ontology modelling and design approach is also context-dependent. Fig.~\ref{fig: modelling patterns} shows an overview of the building ontology modelling patterns.

\begin{enumerate}[wide, itemsep=0pt, topsep=0pt, labelindent=0pt, label=(\roman*)]

\item Haystack and Brick are modelled as ontologies with wider class hierarchies, while RECore and DB are modelled as ontologies with deeper class hierarchies. An ontology with a wider class hierarchy usually has a relatively large number of direct subclasses with a higher Relationship Richness (RROnto). This kind of ontology usually has high concept coverage and compatibility (i.e., constructional efficiency), but keeping the consistency of axioms defined in multiple classes is problematic and affects its functional adequacy. Conversely, an ontology with a deeper class hierarchy is an ontology where classes have a relatively small number of direct subclasses, use properties and attributes more often, and have a higher Properties Richness (PROnto). An ontology with a deeper class hierarchy has low coupling and cohesion (i.e., functional efficiency), but this may affect the discoverability of the classes, and it is not easy to maintain and exchange the data with other ontologies in such an ontology design.

\item RECore and Brick are aimed towards specific detail modelling, while Haystack and DB are aimed towards generic modelling. Generic modelling with no detailed information is more flexible in composing queries and can have small effects on reducing the query execution time. Still, it encapsulates or implies the properties with idioms and usually needs to be supplemented with specific ontologies to describe more specific features where necessary. For example, generic classes can be supplemented with additional metadata on instances to embody information. On the other hand, specific modelling with detailed information aids in discoverability and ease of use for modellers. It is easier to classify an object type if there is already a suitable predefined class, though this can mean more work to maintain such a complex and specific ontology.

\item Based on the findings from our evaluation, we also abstract the key concepts and relations that are used in building ontologies. The overall picture of the ontology design patterns (ODPs) for smart building applications and how these building concepts are linked is shown in Fig.~\ref{fig: design patterns}. The measurements (Status, Substance, and Unit) are placed in the centroid of the diagram, surrounded by Building Spaces, Building Equipment and Systems, and Building Points. The measurements of building equipment and systems include their status (\textit{hasStatus}), and input and output substances (\textit{hasInputSubstance} and \textit{hasOutputSubstance}). We use \textit{hasStatus}, \textit{hasQuantityKind}, and \textit{hasUnit} to describe the sensory point measurements of the building points. There are two relations between building spaces and building equipment and systems, representing the physical location (\textit{hasLocation}) and functional location (\textit{feeds}) of the equipment, respectively. Building points can be mounted on equipment and systems or directly on the building spaces using \textit{hasLocation}. They are defined as a point or function (\textit{hasPoint}/\textit{hasCapability}) provided by building entities (spaces, equipment, and systems).

\begin{figure}[!hbt]
\centering
\includegraphics[width=1\linewidth]{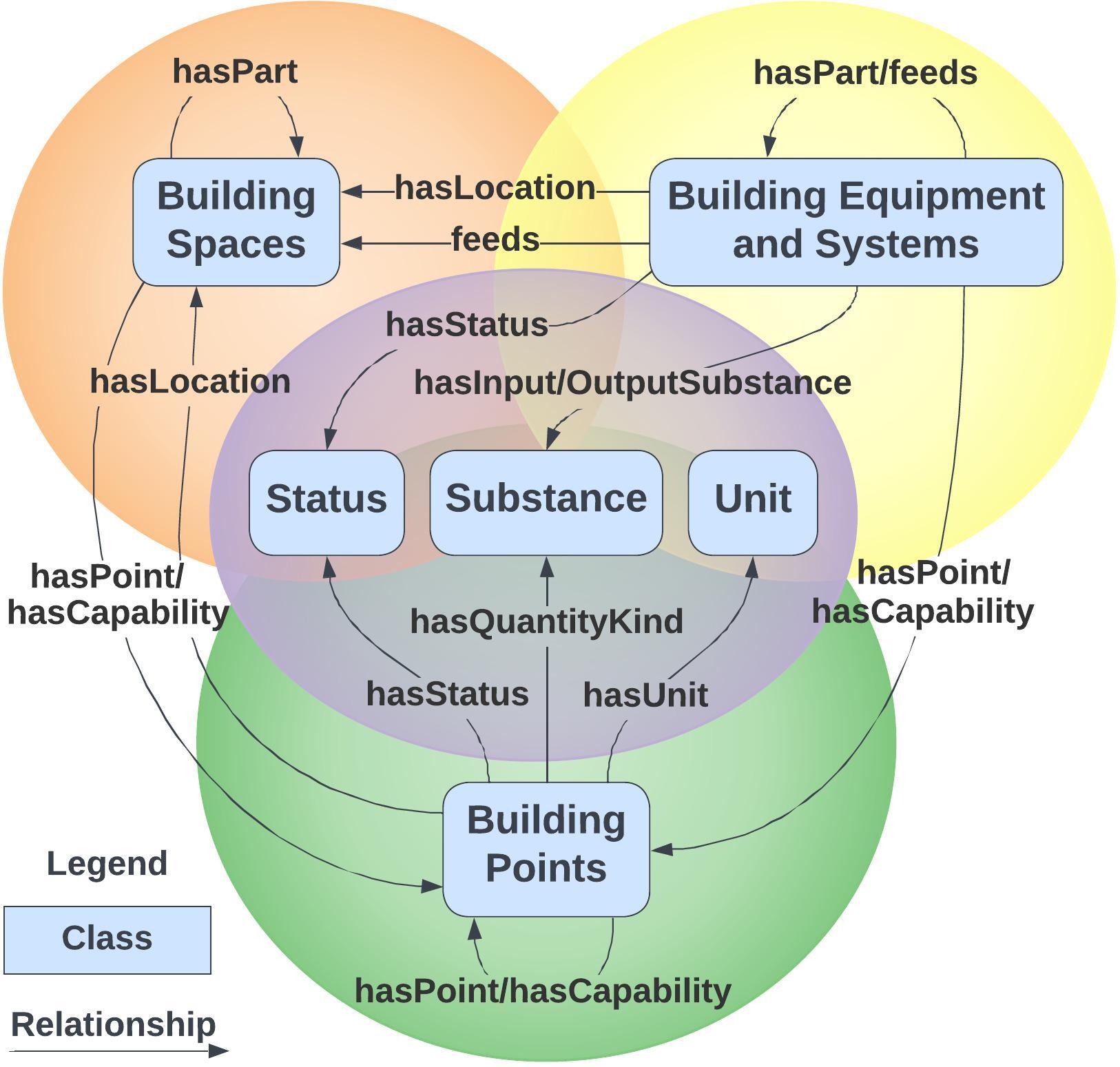}
\caption{The overview of building ontology design patterns.}
\label{fig: design patterns}
\end{figure}

\end{enumerate}

\section{Preliminary Results of Building ODPs}
\label{sec: building ODPs}

Ontology Design Patterns (ODPs) were initially introduced in~\cite{gangemi2009ontology}, aiming to simplify ontology engineering by packaging archetypal solutions into reusable and portable building blocks. The notion of ODPs is to alter ontology engineering to be a composition task with minimal modification, i.e., assembling a collection of reusable ODP implementations and catering to specifics. In this setting, ODPs secure a level of interoperability where data can be integrated. Ontologies that follow the same ODPs have common design choices and inherent points of alignment~\cite{hitzler2016ontology}. Currently, there is a lack of criteria for how existing building ontologies can be reused and re-engineered. The conflicting design rationales, produce a heavy alignment workload to transfer data built by one building ontology to another. ODPs are neither available to the extent required nor are those that are available realised in practice within the smart building domain. Thus, along with this work, we also show some preliminary results of building ODPs that mediate between use cases and design solutions.

\begin{enumerate}[wide, itemsep=0pt, topsep=0pt, labelindent=0pt, label=(\roman*)]

\item \textbf{Building Spaces} 

Fig.~\ref{fig: design pattern building spaces} shows the ODPs for building spaces. The concepts of building spaces include \textit{Region}, \textit{Site}, \textit{Building}, \textit{Building\_Related}, \textit{Floor}, \textit{Zone}, and \textit{Room}. They are related through part-whole relationships. In an open world assumption~\cite{drummond2006open}, it is impossible to enumerate all types of rooms. Therefore, we use a \textit{Category} class to represent the room functions instead. All these classes are categorised into the \textit{Building\_Space} class. The \textit{Building\_Space} class has two properties: \textit{hasGeometry} and \textit{hasTopologicalRelations}, representing the geometry and topological connections, respectively. The sub-properties of \textit{hasTopologicalRelations} are retrieved and evolved from BOT~\cite{rasmussen2021bot} and GeoSPARQL~\cite{battle2011geosparql}.

\begin{figure}[!hbt]
\centering
\includegraphics[width=\linewidth]{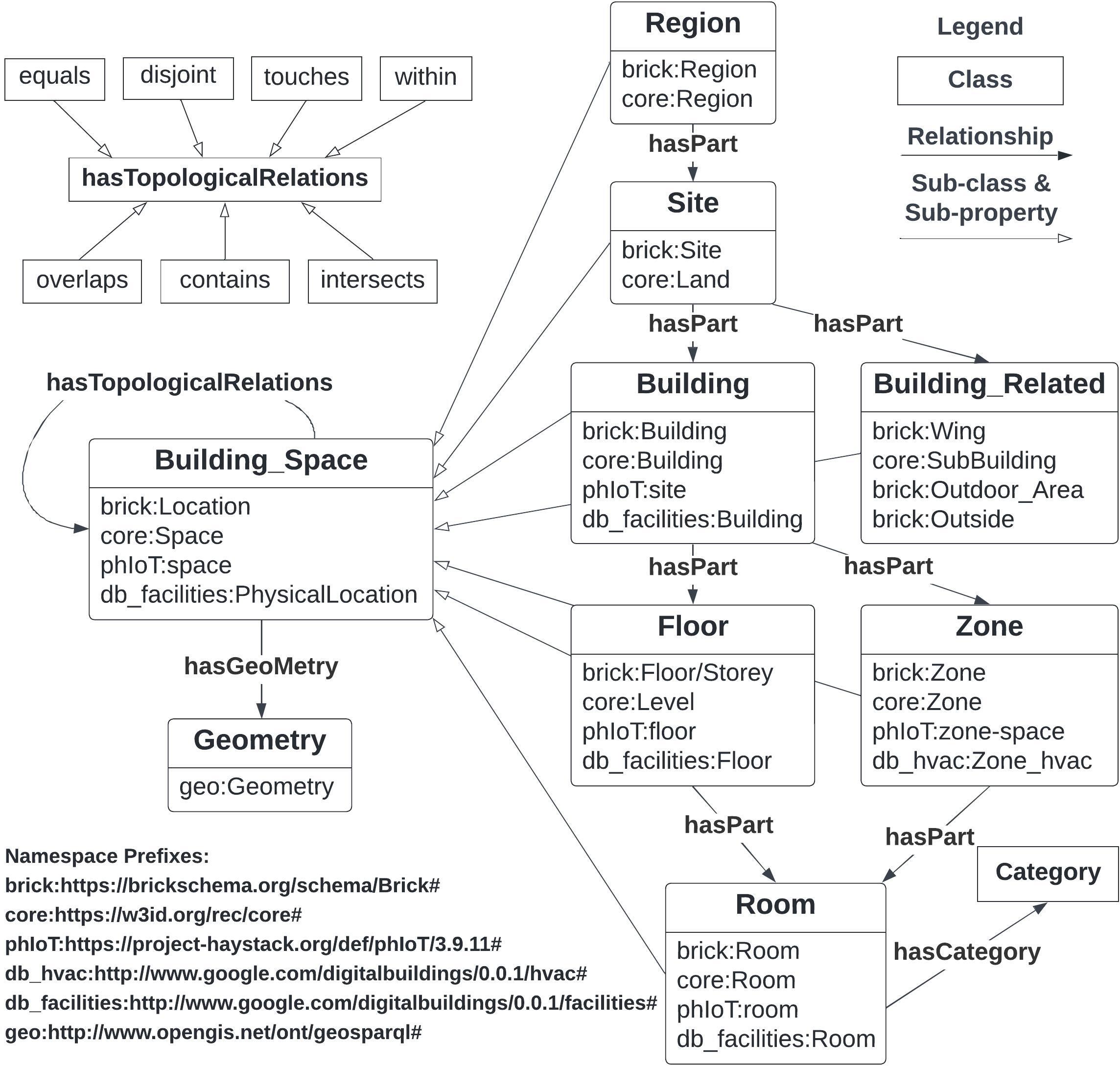}
\caption{ODPs for building spaces.}
\label{fig: design pattern building spaces}
\end{figure}

\item \textbf{Building Equipment and Systems}

Fig.~\ref{fig: design pattern building equipment and systems} shows the ODPs for building equipment and systems. The concepts of building equipment and systems include \textit{Equipment}, \textit{Component}, \textit{System}, and \textit{Loop}. They are related through part-whole relationships and grouped into the \textit{Building\_Equipment\_And\_System} class. \textit{Equipment} has two properties linked to \textit{Building\_Space}, \textit{hasLocation} and \textit{feeds}. They are used to describe the physical and functional location, respectively. \textit{Equipment} and \textit{Component} have a self-contained feeds relationship, used to describe the physical connection and substance transmission between equipment and equipment, or between component and component. The property \textit{feeds} has two sub-properties: \textit{feedsAir} and \textit{feedsWater}. \textit{feedsWater} is detailed into \textit{feedsHotWater} and \textit{feedsChilledWater}. It can be extended to other substances, such as gas or other liquids. \textit{Substance} is linked to the \textit{Equipment} and \textit{Component} by using the property \textit{hasSubstance}. There are two sub-properties \textit{hasInputSubstance} and \textit{hasOutputSubstance} under \textit{hasSubstance}. We suggest the subclasses under \textit{Substance} reuse QUDT~\cite{hodgson2011qudt} and its qualities.

\begin{figure}[!hbt]
\centering
\includegraphics[width=\linewidth]{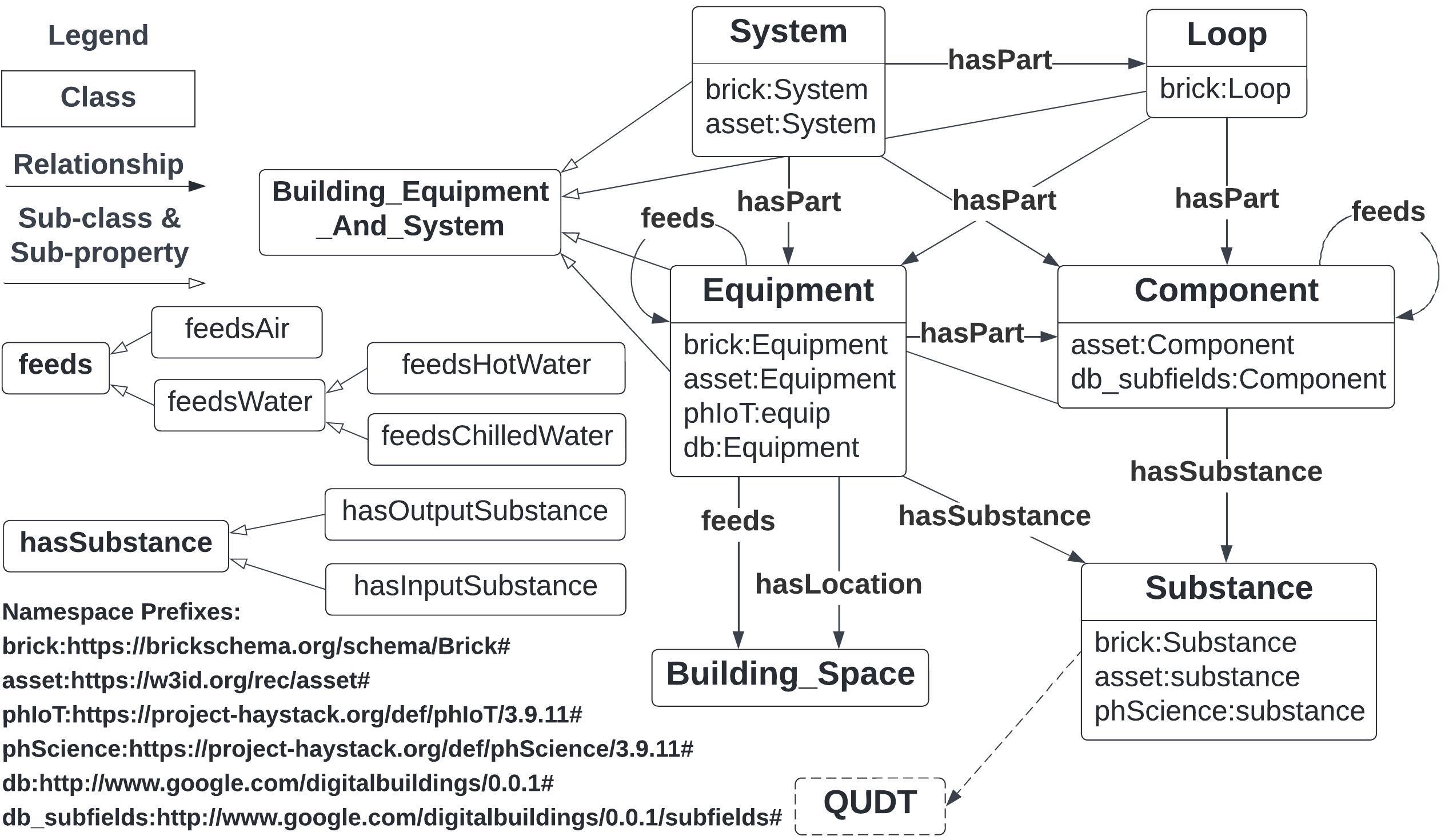}
\caption{ODPs for building equipment and systems.}
\label{fig: design pattern building equipment and systems}
\end{figure}

\item \textbf{Building Points and Measurements}

Fig.~\ref{fig: design pattern building points and measurements} shows the ODPs for building points and measurements. Due to conflicts in defining the building points and properties used, we do not specify the type of the building points, and the properties used to describe the measurement are based on SSN~\cite{compton2012ssn} and SOSA~\cite{haller2019modular}. The property alignment of four building ontologies are listed in Table~\ref{tab: property alignment}. We suggest the subclasses under Substance reuse QUDT~\cite{hodgson2011qudt} and its units.

\begin{figure}[!hbt]
\centering
\includegraphics[width=\linewidth]{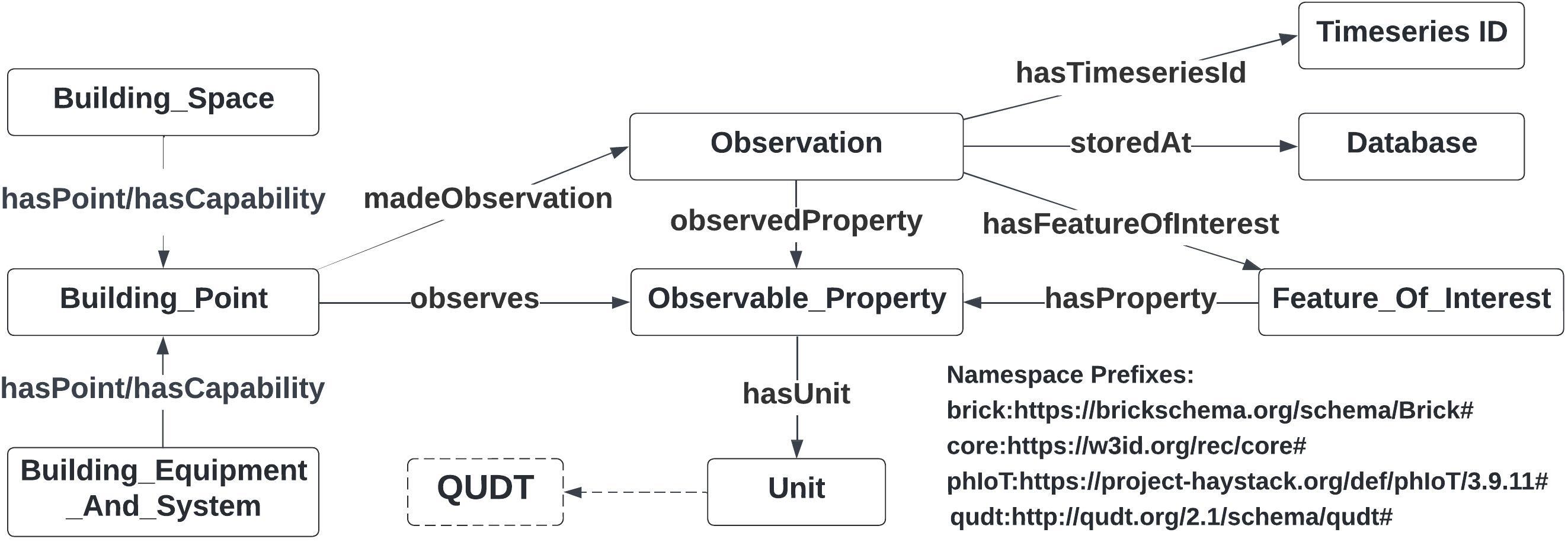}
\caption{ODPs for building points and measurements.}
\label{fig: design pattern building points and measurements}
\end{figure}

\begin{table}[!hbt]
\caption{Property alignment of ODPs for building points and measurements.}
\label{tab: property alignment}
\begin{center}
\begin{adjustbox}{width=1\columnwidth,center}
\begin{tabular}{|l|l|l|l|l|}
\hline
ODPs                       & Brick                          & RECore                    & Haystack              & DB\\  \hline
\small madeObservation     & brick:timeseries\footnotemark  & core:generatedObservation & N/A                   & N/A\\ \hline
\small observes            & brick:measures                 & core:observes             & phIoT:pointQuantity   & N/A\\ \hline
\small hasFeatureOfInterst & brick:measures                 & core:observes             & phIoT:pointSubject    & N/A\\ \hline
\small observedProperty    & N/A                            & qudt:hasQuantityKind      & N/A                   & N/A\\ \hline
\small hasProperty         & N/A                            & N/A                       & N/A                   & N/A\\ \hline
\end{tabular}
\end{adjustbox}
\end{center}
\end{table}

\footnotetext{There is a slight difference between brick:timeseries and madeObservation/core:generatedObservation. brick:timeseries originally points to a time series reference and indicates where and how the data for this point is stored~\cite{balaji2016brick}. It can contain multiple observations for a particular period of time. However, madeObservation/core:generatedObservation only refers to one observation for a particular period of time.}

\end{enumerate}

\section{Conclusions}
\label{sec: conclusions}

This paper compares and evaluates building ontologies for deploying data-driven analytics in smart buildings. The TBox and ABox characteristics of four major building ontologies are compared and evaluated. We also systematically analyse the compatibility between ontologies, and the common and conflicting ontology modelling and design patterns discovered from the study. Since the Brick, RECore, and Haystack designers now plan to align their future developments, we hope the present work will be useful to the authors of these ontologies.

While the evaluation conducted in this study is based on an office building, future work will further the comparison with respect to the capability of extension to other building types, such as commercial and residential buildings. We also plan to apply the comparison and evaluation to more building ontologies to discover solid and generic ontology modelling and design patterns, and to automate the building ontology matching, alignment, and harmonisation. The Linked Building Data Community Group (LBD-CG)~\cite{lbd2023} has similar aims.

We find that there is no ontology in our study that covers all requirements for our use case out-of-the-box. Nor can we universally declare that any ontology is ``better'' than another, even when using popular quantitative metrics. Each ontology has been designed with particular purposes in mind, and when reusing the ontologies, one designed for the purpose closest to that of the new purpose will almost certainly be the best choice~\cite{fierro2022survey}. While efforts to harmonise popular ontologies to expand their expressivity for multiple use cases will help in ontology selection, interoperability, and community knowledge sharing, we believe that a universal building ontology is unreachable. Too much complexity in an ontology can be just as damaging as too little. Even when a certain domain is considered entirely dealt with, there will remain issues with interoperability across multiple domains, for example, across buildings and smart cities. These needs for interoperability will emerge and evolve as the fabric of society evolves, and we must accept that dynamicity is here to stay. In this scenario, the ontology design and modelling patterns we have found in this work can be used as a skeleton to evolve current building ontologies and develop new building ontologies. They can also be used as a guideline to develop alignments for translation between ontologies. We propose to automate the process of pattern discovery as a key element of ontology alignment, integration, and harmonisation activities going forward. The need for these activities will endure and require ongoing research to develop sound and efficient tool support.

\section*{Declaration of Competing Interest}

The authors declare the following financial interests/personal relationships which may be considered as potential competing interests: Kerry Taylor has a patent \url{https://patents.google.com/patent/US9122988B2/en} issued to CSIRO Australia. Kerry Taylor has a patent \url{https://patents.google.com/patent/US8990127/en} issued to CSIRO Australia.

\section*{Data availability}

Data will be made available on request.

\section*{Acknowledgements}

The authors thank the reviewers for their valuable comments to improve the manuscript during the review process. The authors also thank Gregor Henze and Maggie Sullivan from the University of Colorado Boulder for giving helpful advice on the early stages of this work.

This research is a part of the project ``Building and IoT sensor data management'', a cooperative project by the Australian National University (ANU) and the Commonwealth Scientific and Industrial Research Organisation (CSIRO). The authors would like to thank CSIRO for supporting this project.

\section*{Copyright}

© 2023. This manuscript version is made available under the CC-BY-NC-ND 4.0 license~\url{https://creativecommons.org/licenses/by-nc-nd/4.0/}

Contents and figures may change after the correction process. The formal publication is available at:~\url{https://doi.org/10.1016/j.enbuild.2023.113054}

\clearpage
\bibliographystyle{elsarticle-num}
\bibliography{qiang-bibliography-eb}

\appendix

\setcounter{figure}{0}
\setcounter{table}{0}
\renewcommand{\thetable}{A\arabic{table}}

\clearpage
\onecolumn

\section{Building Ontology Corpus}
\label{appendix: building ontology corpus}

\noindent Note: This building ontology corpus is selected and evolved from~\cite{pritoni2021metadata}.

\begin{table}[!hbt]
\begin{adjustbox}{width=1\columnwidth,center}
\begin{tabular}{|l|l|l|}                                                                                                           \hline
No. & Abbreviations & References                                                                                                \\ \hline
\#1   & DogOnt        & Dogont-ontology modeling for intelligent domotic environments~\cite{bonino2008dogont}                   \\ \hline
\#2   & ThinkHome     & Thinkhome energy efficiency in future smart homes~\cite{reinisch2011thinkhome}                          \\ \hline
\#3   & Basont        & Basont-a modular, adaptive building automation system ontology~\cite{ploennigs2012basont}               \\ \hline
\#4   & BOnSAI        & BOnSAI: a smart building ontology for ambient intelligence~\cite{stavropoulos2012bonsai}                \\ \hline
\#5   & SimModel      & SimModel: A domain data model for whole building energy simulation~\cite{o2011simmodel}                 \\ \hline
\#6   & Km4City       & Km4City Smart City API: an integrated support for mobility services~\cite{nesi2016km4city}              \\ \hline
\#7   & SAREF         & Created in close interaction with the industry: the smart appliances reference (SAREF)    
                      ontology~\cite{daniele2015created}                                                                        \\ \hline
\#8   & HTO           & An ontology design pattern for iot device tagging systems~\cite{charpenay2015ontology}                  \\ \hline
\#9   & Brick         & Brick: Towards a unified metadata schema for buildings~\cite{balaji2016brick}                           \\ \hline
\#10  & ifcOWL        & Converting the industry foundation classes to the web ontology language~\cite{schevers2005converting}   \\ \hline
\#11  & oneM2M        & Recent advancements in the Internet-of-Things related standards: A oneM2M perspective~\cite{park2016recent} 
                    \\ \hline
\#12  & SBMS          & Semantic BMS: Ontology for Analysis of Building Operation Efficiency~\cite{kuvcera2017semantic}         \\ \hline
\#13  & WoT           & Introducing Thing Descriptions and Interactions: An Ontology for the Web of Things~\cite{charpenay2016introducing}                     \\ \hline
\#14  & BACS          & Reusing Domain Ontologies in Linked Building Data: the Case of Building Automation and 
                    Control~\cite{terkaj2017reusing}     \\ \hline
\#15  & CRTLont       & Ontology-based modeling of control logic in building automation systems~\cite{schneider2017ontology}    \\ \hline
\#16  & RECore        & The realestatecore ontology~\cite{hammar2019realestatecore}                                             \\ \hline
\#17  & SAREF4BLDG    & Extending the SAREF ontology for building devices and topology~\cite{poveda2018extending}               \\ \hline
\#18  & SBOnto        & SBOnto: Ontology of smart building~\cite{vzavcek2017sbonto}                                             \\ \hline
\#19  & SEAS          & The SEAS Knowledge Model~\cite{lefranccois2017seas}                                                     \\ \hline
\#20  & SSN/SOSA      & Semantic Sensor Network Ontology~\cite{neuhaus2009semantic}                                             \\ \hline
\#21  & Onto-SB       & Onto-SB: human profile ontology for energy efficiency in smart building~\cite{degha2018onto}            \\ \hline
\#22  & OPM           & Ontology for Property Management~\cite{opm2018}                                                         \\ \hline
\#23  & BOT           & BOT: the building topology ontology of the W3C linked building data group~\cite{rasmussen2021bot}       \\ \hline
\#24  & EM-KPI        & Enhancing energy management at district and building levels via an EM-KPI ontology~\cite{yehongkpi}     \\ \hline
\#25  & OnCom         & An Ontology-Based Thermal Comfort Management System In Smart Buildings~\cite{orozco2019ontology}        \\ \hline
\#26  & Haystack      & Project Haystack Data Standards~\cite{john2020project}                                                  \\ \hline
\#27  & DB            & A Digital Buildings Ontology for Google's Real Estate~\cite{berkoben2020digital}                        \\ \hline
\#28  & OP            & From obXML to the OP ontology: developing a semantic model for occupancy profile~\cite{chavez2020obxml} \\ \hline
\#29  & RESPONSE      & Integrating building and iot data in demand response solutions~\cite{esnaola2019integrating}            \\ \hline
\#30  & TUBES         & TUBES System Ontology: Digitalization of building service systems~\cite{pauen2021tubes}                 \\ \hline
\end{tabular}
\end{adjustbox}
\end{table}

\section{Use Case Evaluation of Energy Audits}
\label{appendix: competency questions in energy audits}

\noindent{Note: We used the OWL inferencing engine to execute the queries on our building sample. If the inferencing engine is not used, all the nested relationships should be specified in the SPARQL queries. For example, replacing \textit{rdf:type} with \textit{rdf:type/rdfs:subClassOf*}.}

\begin{center}

\begin{table}[!hbt]
\label{tab: prefix and namespaces}
\begin{center}
\begin{adjustbox}{width=0.8\columnwidth,center}
\begin{tabular}{|l|l|l|}
\hline
Ontologies                          & Prefixes              & Namespaces\\
\hline
\small Brick                        & \small brick          & \small \url{https://brickschema.org/schema/Brick#}\\
\hline
\multirow{4}{*}{\small RECore}      & \small core           & \small \url{https://w3id.org/rec/core#}\\
                                    & \small asset          & \small \url{https://w3id.org/rec/asset#} \\
                                    & \small device         & \small \url{https://w3id.org/rec/device#} \\
                                    & \small analytics      & \small \url{https://w3id.org/rec/analytics#} \\                                 
\hline
\multirow{3}{*}{\small Haystack}    & \small ph             & \small \url{https://project-haystack.org/def/ph/3.9.11#}\\
                                    & \small phIoT          & \small \url{https://project-haystack.org/def/phIoT/3.9.11#}\\
                                    & \small phScience      & \small \url{https://project-haystack.org/def/phScience/3.9.11#}\\
\hline
\multirow{3}{*}{\small DB}          & \small db             & \small \url{http://www.google.com/digitalbuildings/0.0.1#}\\
                                    & \small db\_hvac       & \small \url{http://www.google.com/digitalbuildings/0.0.1/hvac#} \\
                                    & \small db\_subfields  & \small \url{http://www.google.com/digitalbuildings/0.0.1/subfields#} \\
\hline
\end{tabular}
\end{adjustbox}
\end{center}
\end{table}

\begin{longtable}{|M{1.5cm}|m{9cm}|m{6.5cm}|}
\multicolumn{3}{p{17cm}}{(i) Competency Questions Related to Building Spaces:} \\
\multicolumn{3}{p{17cm}}{} \\
\multicolumn{3}{p{17cm}}{\#01: What is the primary use of the building? If multiple uses, what is the fraction of floor area for each use?} \\
\multicolumn{3}{p{17cm}}{}                                                                                  \\ \hline
Ontologies & SPARQL Queries & Query Evaluations                                                             \\ \hline
Brick  & \begin{lstlisting}
SELECT ?function
WHERE {
BLDG:Building_003 brick:buildingPrimaryFunction ?function.
}
\end{lstlisting} & Partially Answered \newline (Can answer the question ``What is the primary use of the building?'' and will return multiple results if the model has multiple uses. Cannot answer ``What is the fraction of floor area for each use?'' because the function is defined for the building, not other location types) \\ \hline
RECore  & \begin{lstlisting} 
SELECT ?function
WHERE {
BLDG:Building_003 core:premisesType ?function.
}\end{lstlisting}  & Partially Answered \newline (Can answer the question ``What is the primary use of the building?'' and will return multiple results if the model has multiple uses. Cannot answer ``What is the fraction of floor area for each use?'' because core:hasPremiseType can be used for administrative grouping of spaces, such as building components, but the definition of floor area is missing)  \\ \hline
Haystack  & \begin{lstlisting} 
SELECT ?function
WHERE {
BLDG:Building_003 phIoT:primaryFunction ?function.
}\end{lstlisting}  & Partially Answered \newline (Can answer the question ``What is the primary use of the building?'' and will return multiple results if the model has multiple uses. Cannot answer ``What is the fraction of floor area for each use?'' because the function is defined for the building, not other location types)   \\ \hline
DB  & N/A   & Unable to answer \newline (No definition of building function type)     \\ \hline

\multicolumn{3}{p{17cm}}{} \\
\multicolumn{3}{p{17cm}}{\#02: What are the HVAC zones and their respective air-conditioned areas in the building?} \\
\multicolumn{3}{p{17cm}}{} \\ \hline
Ontologies & SPARQL Queries & Query Evaluations            \\ \hline
Brick  & \begin{lstlisting} 
SELECT ?HVACZone ?conditionedArea
WHERE {
?HVACZone rdf:type brick:HVAC_Zone.
BLDG:Building_003 brick:hasPart ?HVACZone.
?HVACZone brick:hasPart ?conditionedArea.
}
\end{lstlisting}  & Fully Answered \\ \hline
RECore  & \begin{lstlisting}
SELECT ?HVACZone ?conditionedArea
WHERE {
?HVACZone rdf:type core:HVACZone.
BLDG:Building_003 core:hasPart ?HVACZone.
?HVACZone core:hasPart ?conditionedArea.
}
\end{lstlisting} & Fully Answered \\ \hline
Haystack  & \begin{lstlisting}
SELECT ?HVACZone ?conditionedArea
WHERE {
?HVACZone rdf:type phIoT:hvac-zone-space.
?HVACZone phIoT:siteRef BLDG:Building_003.
?conditionedArea phIoT:spaceRef ?HVACZone.
}
\end{lstlisting} & Fully Answered  \\ \hline
DB  & \begin{lstlisting}
SELECT ?HVACZone ?conditionedArea
WHERE {
?HVACZone rdf:type db_hvac:Zone_hvac.
BLDG:Building_003 db:hasPhysicalLocation ?HVACZone.
?HVACZone db:hasPhysicalLocation ?conditionedArea.
}
\end{lstlisting} & Partially Answered \newline (Can be queried, but using the property \textit{db:hasPhysicalLocation} for HVAC zones is not appropriate) \\ \hline

\multicolumn{3}{p{17cm}}{} \\
\multicolumn{3}{p{17cm}}{\#03: What are all the rooms and their floors in the building?} \\
\multicolumn{3}{p{17cm}}{} \\ \hline
Ontologies & SPARQL Queries & Query Evaluations            \\ \hline
Brick  & \begin{lstlisting} 
SELECT ?floor ?room
WHERE {
?floor rdf:type brick:Floor.
BLDG:Building_003 brick:hasPart+ ?floor.
OPTIONAL {
    ?floor brick:hasPart ?room.
    ?room rdf:type brick:Room.
    }
}
\end{lstlisting}  & Fully Answered \newline (Floor might be part of the building, but it might also be part of a wing that is part of the building. For this reason, we use the ``+'' property path quantifier to say that it is related by at least one step of this property)\\ \hline
RECore  & \begin{lstlisting} 
SELECT ?floor ?room
WHERE {
?floor rdf:type core:Level. 
BLDG:Building_003 core:hasPart+ ?floor.
OPTIONAL {
    ?floor core:hasPart ?room.
    ?room rdf:type core:Room.
    }
}
\end{lstlisting} & Fully Answered \newline (Floor might be part of the building, but it might also be part of a wing that is part of the building. For this reason, we use the ``+'' property path quantifier to say that it is related by at least one step of this property) \\ \hline
Haystack  & \begin{lstlisting}
SELECT ?floor ?room
WHERE {
?floor rdf:type phIoT:floor. 
?floor phIoT:siteRef BLDG:Building_003.
OPTIONAL {
    ?room phIoT:spaceRef ?floor.
    ?room rdf:type phIoT:room.
    }
}
\end{lstlisting} & Fully Answered  \\ \hline
DB  & \begin{lstlisting}
SELECT ?floor ?room
WHERE {
BLDG:Building_003 db:hasFloor ?floor.
OPTIONAL {
    ?floor db:hasRoom ?room.
    }
}
\end{lstlisting} & Partially Answered \newline (Floor might be part of the building, but it might also be part of a wing that is part of the building. As hasFloor is not a transitive property, only non-transitive information can be queried) \\ \hline

\multicolumn{3}{p{17cm}}{} \\
\multicolumn{3}{p{17cm}}{\#04: What is the function type of each room?} \\ 
\multicolumn{3}{p{17cm}}{} \\ \hline
Ontologies & SPARQL Queries & Query Evaluations            \\ \hline
Brick  & \begin{lstlisting} 
SELECT ?room ?roomType
WHERE {
?room rdf:type brick:Room.
BLDG:Building_003 brick:hasPart+ ?room.
?room rdf:type ?roomType.
}
\end{lstlisting} & Fully Answered \\ \hline
RECore  & \begin{lstlisting} 
SELECT ?room ?roomType
WHERE {
?room rdf:type core:Room.
BLDG:Building_003 core:hasPart+ ?room.
?room rdf:type ?roomType.
}
\end{lstlisting} & Fully Answered \\ \hline
Haystack  & \begin{lstlisting}
SELECT ?room ?roomType
WHERE {
?room rdf:type phIoT:room.
?room phIoT:siteRef BLDG:Building_003.
?room rdf:type ?roomType.
}
\end{lstlisting} & Partially Answered \newline (It can be queried, but all rooms are of the same general type) \\ \hline
DB  & \begin{lstlisting}
SELECT ?room ?roomType
WHERE {
BLDG:Building_003 db:hasFloor ?floor.
?floor db:hasRoom ?room.
?room rdf:type ?roomType.
}
\end{lstlisting} & Partially Answered \newline (It can be queried, but all rooms are of the same general type) \\ \hline

\multicolumn{3}{p{17cm}}{} \\
\multicolumn{3}{p{17cm}}{\#05: What is the occupancy of each room?} \\ 
\multicolumn{3}{p{17cm}}{} \\ \hline
Ontologies & SPARQL Queries & Query Evaluations            \\ \hline
Brick  & \begin{lstlisting} 
SELECT ?room ?occupancyStatus
WHERE {
?room rdf:type brick:Room.
BLDG:Building_003 brick:hasPart+ ?room.
?room brick:hasPoint ?occupancyStatus. 
?occupancyStatus rdf:type brick:Occupancy_Status.
}
\end{lstlisting} & Fully Answered \\ \hline
RECore & \begin{lstlisting} 
SELECT ?room ?occupancyStatus
WHERE {
?room rdf:type core:Room.
BLDG:Building_003 core:hasPart+ ?room.
?room core:hasCapability ?occupancyStatus. 
?occupancyStatus rdf:type device:OccupancyState.
}
\end{lstlisting} & Fully Answered \\ \hline
Haystack         & \begin{lstlisting}
SELECT ?room ?occupancyStatus
WHERE {
?room rdf:type phIoT:room.
?room phIoT:siteRef BLDG:Building_003.
?occupancyStatus phIoT:spaceRef ?room. 
?occupancyStatus rdf:type phIoT:occupied.
}
\end{lstlisting}  & Fully Answered \\ \hline
DB  & N/A & Unable to Answer \newline (Although DB has db\_subfields:Occupancy to define the status of being occupied, there is no appropriate property that can be found to link db\_subfields:Room and db\_subfields:Occupancy) \\ \hline

\multicolumn{3}{p{17cm}}{} \\
\multicolumn{3}{p{17cm}}{(ii) Competency Questions Related to Building Equipment and Systems:} \\
\multicolumn{3}{p{17cm}}{} \\
\multicolumn{3}{p{17cm}}{\#06: What are the pieces of equipment and systems of equipment in the building?} \\ 
\multicolumn{3}{p{17cm}}{} \\ \hline
Ontologies & SPARQL Queries & Query Evaluations               \\ \hline
Brick  & \begin{lstlisting}
SELECT ?equipment ?system 
WHERE {
{?equipment rdf:type brick:Equipment.}
UNION
{?system rdf:type brick:System.}
}
\end{lstlisting} & Fully Answered \\ \hline
RECore  & \begin{lstlisting}
SELECT ?equipment ?system
WHERE {
{?equipment rdf:type asset:Equipment.}
UNION
{?system rdf:type asset:System.}
}
\end{lstlisting} & Fully Answered \\ \hline
Haystack  & \begin{lstlisting}
SELECT ?equipment
WHERE {
?equipment rdf:type phIoT:equip. 
}
\end{lstlisting} & Partially Answered \newline (The definition of systems of equipment is not fully implemented in the ontology at the time of writing) \\ \hline
DB  & \begin{lstlisting}
SELECT ?equipment
WHERE {
?equipment rdf:type db:Equipment. 
}
\end{lstlisting} & Partially Answered \newline (No definition of systems of equipment) \\ \hline

\multicolumn{3}{p{17cm}}{} \\
\multicolumn{3}{p{17cm}}{\#07: What are the properties (e.g., type, physical location, and functional area) of each piece of HVAC equipment?} \\
\multicolumn{3}{p{17cm}}{} \\ \hline
Ontologies & SPARQL Queries & Query Evaluations               \\ \hline
Brick  & \begin{lstlisting}
SELECT ?equipment ?type ?physicalLocation ?functionalArea
WHERE {
?equipment rdf:type brick:HVAC_Equipment.
?equipment rdf:type ?type.
?equipment brick:hasLocation ?physicalLocation.
?equipment brick:feedsAir+ ?functionalArea.
}
\end{lstlisting} & Fully Answered \\ \hline
RECore  & \begin{lstlisting}
SELECT ?equipment ?type ?physicalLocation ?functionalArea
WHERE {
?equipment rdf:type asset:HVACEquipment. 
?equipment rdf:type ?type.
?equipment core:locatedIn ?physicalLocation.
?equipment asset:feeds+ ?functionalArea.
}
\end{lstlisting} & Fully Answered \\ \hline
Haystack  & \begin{lstlisting}
SELECT ?equipment ?type ?physicalLocation ?functionalArea
WHERE {
?equipment rdf:type phIoT:airHandlingEquip. 
?equipment rdf:type ?type.
?equipment phIoT:spaceRef ?physicalLocation.
?functionalArea phIoT:airRef+ ?equipment.
}
\end{lstlisting} & Fully Answered \\ \hline
DB  & \begin{lstlisting}
SELECT ?equipment ?type ?physicalLocation
WHERE {
?equipment rdf:type db:Equipment. 
?equipment rdf:type ?type.
?equipment db:hasPhysicalLocation ?physicalLocation.
}
\end{lstlisting} & Partially Answered \newline (No appropriate property can be found to link db:Equipment and its functional areas)   \\ \hline

\multicolumn{3}{p{17cm}}{} \\
\multicolumn{3}{p{17cm}}{\#08: What are the components of each piece of HVAC equipment and their structural relations?} \\ 
\multicolumn{3}{p{17cm}}{} \\ \hline
Ontologies & SPARQL Queries & Query Evaluations               \\ \hline
Brick  & \begin{lstlisting}
SELECT ?equipment ?component ?feedsAirComponent
WHERE {
?equipment rdf:type brick:HVAC_Equipment.
?equipment brick:hasPart+ ?component.
OPTIONAL {
    ?component brick:feedsAir ?feedsAirComponent. 
    }
}
\end{lstlisting} & Fully Answered \\ \hline
RECore  & \begin{lstlisting}
SELECT ?equipment ?component ?feedsAirComponent
WHERE {
?equipment rdf:type asset:HVACEquipment.
?equipment core:hasPart+ ?component.
OPTIONAL {
    ?component asset:feeds ?feedsAirComponent. 
    }
}
\end{lstlisting} & Fully Answered \\ \hline
Haystack  & \begin{lstlisting}
SELECT ?equipment ?component ?feedsAirComponent
WHERE {
?equipment rdf:type phIoT:airHandlingEquip.
?component phIoT:equipRef+ ?equipment.
OPTIONAL {
    ?feedsAirComponent phIoT:airRef ?component.
    }
}
\end{lstlisting} & Fully Answered \\ \hline
DB  & N/A & Unable to Answer \newline (The connection properties hasPart and feeds are not fully implemented in the ontology at the time of writing)   \\ \hline

\multicolumn{3}{p{17cm}}{} \\
\multicolumn{3}{p{17cm}}{\#09: What is the energy use broken down by resource using the energy?} \\ 
\multicolumn{3}{p{17cm}}{} \\ \hline
Ontologies & SPARQL Queries & Query Evaluations               \\ \hline
Brick  & \begin{lstlisting}
SELECT ?equipment ?substance ?timeseries
WHERE {
?equipment rdf:type brick:Equipment.
?equipment brick:hasInputSubstance ?substance.
?substance rdf:type brick:Substance.
?equipment brick:hasPoint ?point.
?point rdf:type brick:Energy_Sensor.
?point brick:timeseries ?timeseries.
FILTER (?substance = brick:Chilled_Water)
}
\end{lstlisting} & Fully Answered \newline (Assuming the energy use is represented as a time series associated with the energy sensor. Substance must be specified in the query, e.g., brick:Chilled\_Water) \\ \hline
RECore  & \begin{lstlisting}
SELECT ?equipment ?substance ?timeseries
WHERE {
?equipment rdf:type asset:Equipment.
?equipment asset:substance ?substance.
?equipment core:hasCapability ?point.
?point rdf:type device:EnergySensor.
?point core:generatedObservation ?timeseries.
FILTER (?substance = "ChilledWater")
}
\end{lstlisting} & Partially Answered \newline (Assuming the energy use is represented as a time series associated with the energy sensor. Substance must be specified in the query, e.g., ``ChilledWater''. The property of asset:substance has no distinction between input and output)  \\ \hline
Haystack  & \begin{lstlisting}
SELECT ?equipment ?substance ?value ?dateTime
WHERE {
?equipment rdf:type phIoT:equip.
?equipment ph:inputs ?substance.
?substance rdf:type phScience:substance.
?point phIoT:equipRef ?equipment.
?point phIoT:pointFunction phIoT:sensor.
?point phIoT:curVal ?value.
?value ph:dateTime ?dateTime.
FILTER (?substance = phIoT:chilled-water)
}
\end{lstlisting} & Partially Answered \newline (Substance must be specified in the query, e.g., phIoT:chilled-water. Sensor type cannot be specified for Energy Sensor. Use the current point value with date and time as an alternative way to show the time-series data. Subscription to real-time sensor data is not achievable within the ontology, but may be possible by plugging in a SCRAM-based authentication protocol)  \\ \hline
DB  & N/A & Unable to Answer \newline (No properties related to  substance)  \\ \hline

\multicolumn{3}{p{17cm}}{} \\
\multicolumn{3}{p{17cm}}{\#10: What is the energy use broken down by end-use?} \\ 
\multicolumn{3}{p{17cm}}{} \\ \hline
Ontologies & SPARQL Queries & Query Evaluations               \\ \hline
Brick  & \begin{lstlisting}
SELECT ?equipment ?type ?timeseries
WHERE {
?equipment rdf:type brick:Equipment.
?equipment rdf:type ?type.
?equipment brick:hasPoint ?point.
?point rdf:type brick:Energy_Sensor.
?point brick:timeseries ?timeseries.
FILTER (?type = brick:HVAC_Equipment)
}
\end{lstlisting} & Fully Answered \newline (Assuming the energy use is represented as a time series associated with energy sensor. End-use equipment must be specified in the query, e.g., brick:HVAC\_Equipment)  \\ \hline
RECore  & \begin{lstlisting}
SELECT ?equipment ?type ?timeseries
WHERE {
?equipment rdf:type asset:Equipment.
?equipment rdf:type ?type.
?equipment core:hasCapability ?point.
?point rdf:type device:EnergySensor.
?point core:generatedObservation ?timeseries.
FILTER (?type = asset:HVACEquipment)
}
\end{lstlisting} & Fully Answered \newline (Assuming the energy use is represented as a time series associated with energy sensor. End-use equipment must be specified in the query, e.g., asset:HVACEquipment)  \\ \hline
Haystack  & \begin{lstlisting}
SELECT ?equipment ?type ?value ?dateTime
WHERE {
?equipment rdf:type phIoT:equip.
?equipment rdf:type ?type.
?point phIoT:equipRef ?equipment.
?point phIoT:pointFunction phIoT:sensor.
?point phIoT:curVal ?value.
?value ph:dateTime ?dateTime.
FILTER (?type = phIoT:airHandlingEquip)
}
\end{lstlisting} & Partially Answered \newline (End-use equipment must be specified in the query, e.g., phIoT:airHandlingEquip. Sensor type cannot be specified for Energy Sensor. Use the current point value with date and time as an alternative way to show the time-series data. Subscription to real-time sensor data is not achievable within the ontology, but may be possible by plugging in a SCRAM-based authentication protocol)  \\ \hline
DB  & N/A & Unable to answer \newline (No classification of end-use)  \\ \hline

\multicolumn{3}{p{17cm}}{} \\
\multicolumn{3}{p{17cm}}{\#11: What are the energy-generating pieces of equipment in the building?} \\
\multicolumn{3}{p{17cm}}{} \\ \hline
Ontologies & SPARQL Queries & Query Evaluations               \\ \hline
Brick  & \begin{lstlisting}
SELECT ?equipment ?type
WHERE {
{?equipment rdf:type brick:PV_Panel.}
UNION
{?equipment rdf:type brick:Solar_Thermal_Collector.}
?equipment rdf:type ?type.
}
\end{lstlisting} & Partially Answered \newline (Some classes can be found, but no clear classification. Each relevant class should be specified in the query, e.g., brick:PV\_Panel and brick:Solar\_Thermal\_Collector) \\ \hline
RECore  & \begin{lstlisting}
SELECT ?equipment ?type
WHERE {
?equipment rdf:type asset:Generator.
?equipment rdf:type ?type.
}
\end{lstlisting} & Fully Answered \\ \hline
Haystack  & N/A & Unable to Answer \newline (No definitions of energy-generating pieces of equipment)  \\ \hline
DB  & N/A & Unable to Answer \newline (No definitions of energy-generating pieces of equipment)   \\ \hline

\multicolumn{3}{p{17cm}}{} \\
\multicolumn{3}{p{17cm}}{(iii) Competency Questions Related to Building Points and Measurements:} \\
\multicolumn{3}{p{17cm}}{} \\
\multicolumn{3}{p{17cm}}{\#12: What are the meters and meter types in the building?} \\
\multicolumn{3}{p{17cm}}{} \\ \hline
Ontologies & SPARQL Queries & Query Evaluations               \\ \hline
Brick  & \begin{lstlisting}
SELECT ?meter ?meterType
WHERE {
?meter rdf:type brick:Meter.
?meter rdf:type ?meterType.
}
\end{lstlisting} & Fully Answered \\ \hline
RECore  & \begin{lstlisting}
SELECT ?meter ?meterType
WHERE {
?meter rdf:type asset:Meter.
?meter rdf:type ?meterType.
}
\end{lstlisting} & Fully Answered \\ \hline
Haystack  & \begin{lstlisting}
SELECT ?meter ?meterType
WHERE {
?meter rdf:type phIoT:meter.
?meter rdf:type ?meterType.
}
\end{lstlisting} & Fully Answered  \\ \hline
DB  & N/A &  Unable to Answer \newline (No definition of meter)  \\ \hline

\multicolumn{3}{p{17cm}}{} \\
\multicolumn{3}{p{17cm}}{\#13: What does each meter measure (i.e., substance and quantity)?} \\
\multicolumn{3}{p{17cm}}{} \\ \hline
Ontologies & SPARQL Queries & Query Evaluations               \\ \hline
Brick  & \begin{lstlisting}
SELECT ?meter ?substanceAndQuantitykind
WHERE {
?meter rdf:type brick:Meter.
?meter brick:measures ?substanceAndQuantitykind.
}
\end{lstlisting} & Partially Answered \newline (Properties of substance and quantity exist, but it is not possible to relate the substance measured to the quantity kind, unless it is assumed that each meter measures only one substance)  \\ \hline
RECore  & \begin{lstlisting}
SELECT ?meter ?substanceAndQuantitykind
WHERE {
?meter rdf:type asset:Meter.
?meter core:observes ?substanceAndQuantitykind.
}
\end{lstlisting} & Partially Answered \newline (Properties of substance and quantity exist, but it is not possible to relate the substance measured to the quantity kind, unless it is assumed that each meter measures only one substance) \\ \hline
Haystack  & \begin{lstlisting}
SELECT ?meter ?substance ?quantitykind
WHERE {
?meter rdf:type phIoT:meter.
?meter phIoT:pointSubject ?substance.
?meter phIoT:pointQuantity ?quantitykind.
}
\end{lstlisting} & Fully Answered  \\ \hline
DB  & N/A & Unable to Answer \newline (No definition of meter) \\ \hline

\multicolumn{3}{p{17cm}}{} \\
\multicolumn{3}{p{17cm}}{\#14: What are the connections between meters and sub-meters?} \\
\multicolumn{3}{p{17cm}}{} \\ \hline
Ontologies & SPARQL Queries & Query Evaluations               \\ \hline
Brick  & \begin{lstlisting}
SELECT ?meter ?subMeter
WHERE {
?meter rdf:type brick:Meter.
?subMeter rdf:type brick:Meter.
?meter brick:feeds ?subMeter.
}
\end{lstlisting} & Partially Answered \newline (We adopt brick:feeds as an unofficial convention in our models: meter closest to utility point of connection ``feeds'' sub-meters) \\ \hline
RECore  & \begin{lstlisting}
SELECT ?meter ?subMeter
WHERE {
?meter rdf:type asset:Meter.
?subMeter rdf:type asset:Meter.
?meter core:feeds ?subMeter.
}
\end{lstlisting} & Partially Answered \newline (We adopt core:feeds as an unofficial convention in our models: meter closest to utility point of connection ``feeds'' sub-meters)  \\ \hline
Haystack  & \begin{lstlisting}
SELECT ?meter ?subMeter
WHERE {
?meter rdf:type phIoT:meter.
?subMeter rdf:type phIoT:meter.
?subMeter phIoT:submeterOf ?meter.
}
\end{lstlisting} & Fully Answered  \\ \hline
DB  & N/A & Unable to Answer \newline (No definition of meter)  \\ \hline

\multicolumn{3}{p{17cm}}{} \\
\multicolumn{3}{p{17cm}}{\#15: What are the operating parameters of each piece of equipment (e.g., setpoints and schedules)?} \\
\multicolumn{3}{p{17cm}}{} \\ \hline
Ontologies & SPARQL Queries & Query Evaluations               \\ \hline
Brick  & \begin{lstlisting}
SELECT ?equipment ?parameter
WHERE {
?equipment rdf:type brick:Equipment.
?equipment brick:hasPoint ?parameter.
?parameter rdf:type brick:Point.
}
\end{lstlisting} & Fully Answered \\ \hline
RECore  & \begin{lstlisting}
SELECT ?equipment ?parameter
WHERE {
?equipment rdf:type asset:Equipment.
?equipment core:hasCapability ?parameter.
?parameter rdf:type core:Capability.
}
\end{lstlisting} & Fully Answered  \\ \hline
Haystack  & \begin{lstlisting}
SELECT ?equipment ?parameter
WHERE {
?equipment rdf:type phIoT:equip.
?point phIoT:equipRef ?equipment.
?point rdf:type phIoT:point.
?point phIoT:pointFunction ?parameter.
}
\end{lstlisting} & Fully Answered  \\ \hline
DB  & N/A & Unable to answer \newline (No appropriate property can be found to link db:Equipment and its operating parameters)  \\ \hline

\multicolumn{3}{p{17cm}}{} \\
\multicolumn{3}{p{17cm}}{\#16: What are the time series for each meter?} \\
\multicolumn{3}{p{17cm}}{} \\ \hline
Ontologies & SPARQL Queries & Query Evaluations               \\ \hline
Brick  & \begin{lstlisting}
SELECT ?meter ?timeseriesId
WHERE {
?meter rdf:type brick:Meter.
?meter brick:timeseries ?timeseries.
?timeseries brick:hasTimeseriesId ?timeseriesId.
}
\end{lstlisting} & Fully Answered \\ \hline
RECore  & \begin{lstlisting}
SELECT ?meter ?timeseriesId
WHERE {
?meter rdf:type asset:Meter.
?meter core:generatedObservation ?timeseries.
?timeseriesId analytics:seriesMembers ?timeseries.
}
\end{lstlisting} & Fully Answered  \\ \hline
Haystack  & \begin{lstlisting}
SELECT ?meter ?value ?dateTime
WHERE {
?meter rdf:type phIoT:meter.
?meter phIoT:curVal ?value.
?value ph:dateTime ?dateTime.
}
\end{lstlisting} & Particularly Answered \newline(Use the current point value with date and time as an alternative way to show the time-series data. Subscription to real-time sensor data is not achievable within the ontology, but may be possible by plugging in a SCRAM-based authentication protocol)  \\ \hline
DB  & N/A & Unable to Answer \newline (No definition of meter) \\ \hline

\end{longtable}
\end{center}

\clearpage
\twocolumn

\end{document}